\documentclass[aps,prx,twocolumn,superscriptaddress,showpacs]{revtex4}  % for review and submission

\usepackage{graphicx}  % needed for figures
\usepackage{epstopdf}
\usepackage{dcolumn}   % needed for some tables
\usepackage{bm}        % for math
\usepackage{amsmath}
\usepackage{amssymb}   % for math
\usepackage{array}
\usepackage{pdfsync}

\usepackage[caption=false]{subfig}
\usepackage[bookmarks=false,linkcolor=blue,urlcolor=blue,colorlinks,citecolor=blue]{hyperref}
\usepackage{exscale,relsize}
\usepackage{bbm}
\usepackage{subfig}
\def \ua{{\uparrow}}
\def \da{{\downarrow}}

\def \tx#1{{\rm#1}}

\DeclareMathOperator{\sgn}{sgn}

\begin{document}

\author{Eran Sagi}
\affiliation{Department of Condensed Matter Physics, Weizmann Institute of Science, Rehovot, Israel 76100}

\author{Arbel Haim}
\affiliation{Department of Condensed Matter Physics, Weizmann Institute of Science, Rehovot, Israel 76100}

\author{Erez Berg}
\affiliation{ Department of Physics, University of Chicago, Chicago, Illinois 60637, USA}

\author{Felix von Oppen}
\affiliation{Dahlem Center for Complex Quantum Systems and Fachbereich Physik, Freie Universit¨at Berlin, 14195 Berlin, Germany}

\author{Yuval Oreg}

\affiliation{Department of Condensed Matter Physics, Weizmann Institute of Science, Rehovot, Israel 76100}

\title{Fractional chiral superconductors}
\begin{abstract}

Two-dimensional $p_x+ip_y$ topological superconductors host gapless Majorana edge modes,
as well as Majorana bound states at the core of $h/2e$ vortices. Here
we construct a model realizing the fractional
counterpart of this phase: a fractional chiral superconductor.
Our model is composed of an array of coupled Rashba wires
in the presence of strong interactions, Zeeman field, and proximity
coupling to an $s$-wave superconductor. We define the filling factor as $\nu=l_{\text{so}}n/4$, where $n$ is the electronic density and $l_{\text{so}}$ is the spin-orbit length. Focusing on filling $\nu=1/m$, with $m$ being an odd integer, we obtain
a tractable model which allows us to study the properties of the bulk and the edge. Using an $\epsilon$-expansion with $m=2+\epsilon$, we show that the bulk Hamiltonian is gapped and that the edge of the sample hosts a chiral $\mathbb{Z}_{2m}$ parafermion theory with central charge $c=\frac{2m-1}{m+1}$. The tunneling density
of states associated with this edge theory exhibits an anomalous energy dependence of the form $\omega^{m-1}$. Additionally,
we show that $\mathbb{Z}_{2m}$ parafermionic bound states reside at the cores of
$h/2e$ vortices.
Upon
constructing an appropriate Josephson junction in
our system, we find that the
current-phase relation
displays a $4\pi m$
periodicity, reflecting the underlying non-abelian excitations.
\end{abstract}
%anyons (electronic structure), fractional statistics system, Anyons (nonconventional mechanisms in superconductivity)
\maketitle
\emph{Introduction.---}
Much is nowadays known
about topological states which can be understood from a non-interacting
limit. In this limit, the topological states protected by time-reversal,
particle-hole, and chiral symmetries have been classified \cite{Schnyder2008,Kitaev2009}. Beyond this celebrated periodic table, topological
phases protected by other symmetries such as lattice symmetries \cite{Fu2011,Hsieh2012,Tanaka2012,Dziawa2012, Xu2012,Liu2014,Fang2015,Shiozaki2015,Wang2016}, have also been thoroughly studied.

In contrast, strongly interacting topologically ordered phases which cannot be understood in the absence of interactions,
have yet to be fully understood and classified. Since interactions stabilize these phases, their study generally requires non-perturbative
methods. A relatively recent approach to this problem is the coupled-wires
approach \cite{Kane2002,Teo2014,Klinovaja2013c,Seroussi2014,Neupert2014,Sagi2014,Klinovaja2014a,Meng2014,Santos2015,Sagi2015a,Gorohovsky2015,Meng2015,Mross2015,Meng2015a,Sagi2015,Meng2016,Isobe2015,Sahoo2015,Iadecola2016,Fuji2016,Huang2016},
in which a two-dimensional (2D) or three-dimensional (3D) topologically ordered phase is
constructed from an array of coupled quantum wires. The available
machinery for studying interacting one-dimensional (1D) systems plays
a central role in allowing one to write exactly solvable or analytically tractable model systems realizing such non-trivial
phases of matter.

In this work we employ this approach to study \emph{fractional} chiral
superconductors (FCSC) in 2D.
FCSC phases were previously introduced, mostly from
a field-theoretic point of view, in Refs. \cite{Vaezi2013,Mong2014,Vaezi2014}.
The coupled-wires approach provides us with a tractable model which admits direct study of the properties of such phases.

We begin by constructing a non-interacting $p_x+ip_y$ topological superconductor
(TSC). Following Ref. \cite{Seroussi2014}, this is done by placing an array of $N$ non-interacting weakly-coupled Rashba wires in proximity to an $s$-wave superconductor and applying a Zeeman field.
We
then add the effects of interactions and
show that a fractional chiral superconductor (FCSC) can emerge
at fractional filling factors, defined through the spin-orbit length (see Eq. \ref{eq:filling}).  The main observation that will allow
us to construct a model realizing the strongly-interacting FCSC phase is that, at fractional filling factors, one can define new local fermionic operators, composed of an electron dressed by a particle-hole excitation, for which the
effective filling factor becomes an integer. Exploiting this fact, we will
use the results of Ref. \cite{Seroussi2014} to write a $p_x+ip_y$ topological
superconductor in terms of the transformed fermions. Remarkably, in terms of the original fermions the resulting phase is a FCSC.

\begin{figure}
\begin{tabular}{lr}
\hskip -3mm
\rlap{\parbox[c]{0cm}{\vspace{0cm}\footnotesize{(a)}}}
\includegraphics[clip=true,trim =0cm -1cm 0cm 0cm,width=0.25\textwidth]{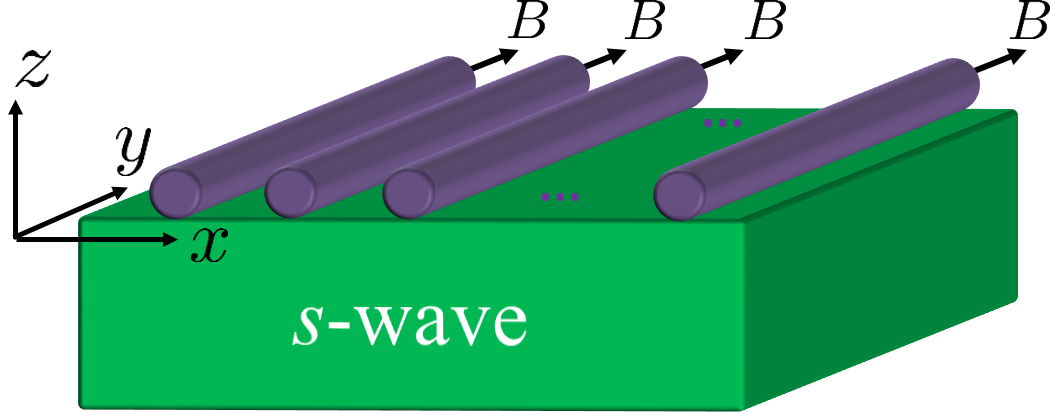} &
\rlap{\parbox[c]{0cm}{\vspace{0cm}\footnotesize{(b)}}}
\includegraphics[clip=true,trim =0cm 0cm 0cm 0cm,width=0.2\textwidth]{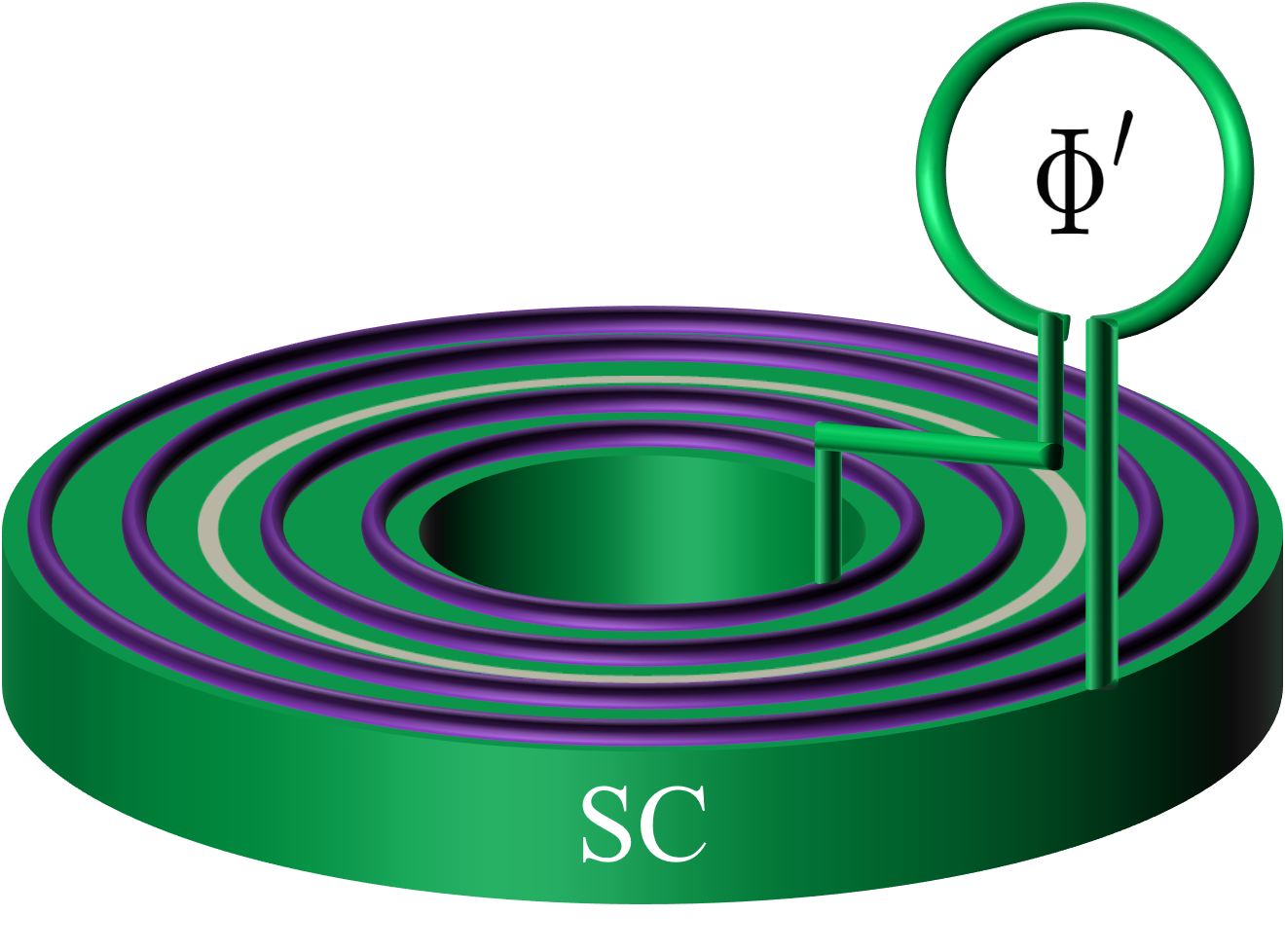}
\end{tabular}
\protect\caption{\label{fig:system} (a) A schematic view of the system we use to construct
a two-dimensional fractional chiral superconductor (FCSC). The
system is composed of an array of $N$ weakly coupled wires with strong
Rashba spin-orbit coupling. We demonstrate that the interplay of the
Zeeman field, proximity to an $s$-wave superconductor, and strong
interactions can result in FCSC phase with a chiral parafermion CFT
at the edges of the sample and parafermionic bound states at
the core of vortices. (b) A corbino geometry, used in order to study $h/2e$ vortices and the anomalous Josephson effect. By connecting the internal and external parts with a superconducting wire, through which flux
can be inserted, the relative phases between the two superconducting
regions can be controlled. It is argued that in the quasi-one-dimensional regime this configuration is equivalent to the effective junction shown in Fig. \ref{fig:SFS}, leading to a $4\pi m$-periodic Josephson effect.}
\end{figure}

After constructing the model, we analyze it by mapping the low-energy theory of the individual wires at fractional fillings $\nu=1/m$ with odd $m$ to a generalized Sine-Gordon model. Performing an $\epsilon$-expansion
 for $m=2+\epsilon$, we show that each wire can be tuned to a $\mathbb{Z}_{2m}$ parafermion multicritical point \cite{zamolodchikov1985}. We then demonstrate that one can couple different wires in such a way that a chiral $\mathbb{Z}_{2m}$ parafermion conformal field theory (CFT) is left on the edge (with $m$ being an odd integer, i.e., $\mathbb{Z}_2, \mathbb{Z}_6, \mathbb{Z}_{10},\cdots$ theories). By construction, the resulting 2D model is topologically ordered.

\begin{figure*}
\subfloat[\label{fig:annulus}]{\includegraphics[scale=0.53]{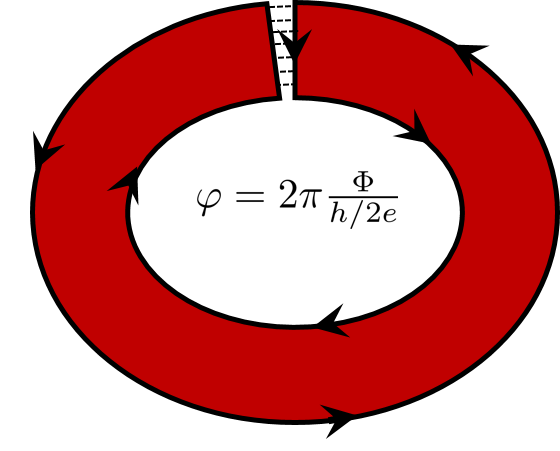}

}\subfloat[\label{fig:bone shaped configuration}]{\includegraphics[scale=0.51]{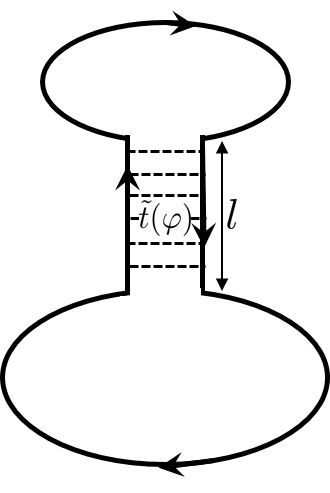}

}\subfloat[\label{fig:shrinking}]{\includegraphics[scale=0.55]{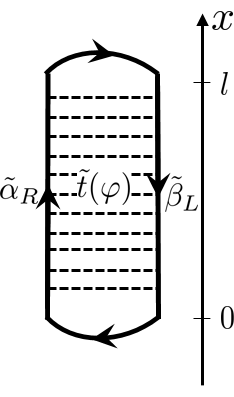}

}\subfloat[\label{fig:auxiliary model}]{\includegraphics[scale=0.53]{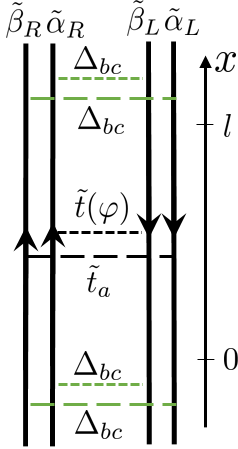}

}\subfloat[\label{fig:SFS-1}]{\includegraphics[scale=0.53]{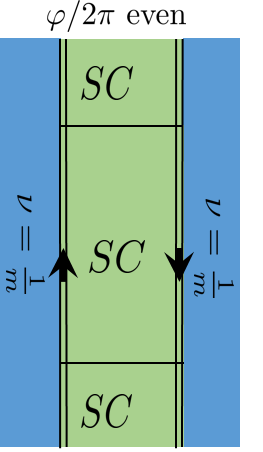}

}\subfloat[\label{fig:SFS}]{\includegraphics[scale=0.53]{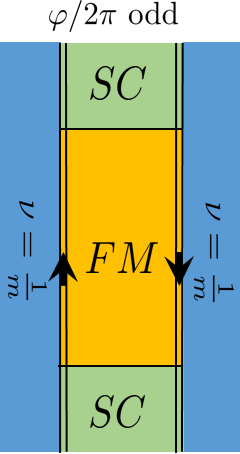}

}

\protect\caption{A schematic depiction of the line of arguments used in order to identify
the zero-modes localized at the center of $h/2e$ vortices. (a) We
start by placing our FCSC phase in an annular geometry, through which
a magnetic flux may be inserted. Without altering the essential
characteristics of the system, we cut a thin trench through the annulus.
(b) As the presence of zero-modes should not depend on the precise
geometry as long as the topological structure is preserved, we focus
on the edge theory and fold our system inside-out to a bone shaped geometry.
(c) The independence on the precise geometry allows us to shrink the
outermost regions. (d) We examine an auxiliary model which has twice as
many degrees of freedom, but whose low energy sector coincides with
our model. We then show that the auxiliary model in (d) can be adiabatically
 deformed into a system composed of $\nu=1/m$ fractional quantum
Hall edges coupled by ferromagnetic and superconducting terms. (e)
If the flux $\varphi$ is an even multiple of $2\pi$, we get effective
superconducting term throughout the system, meaning there are no
zero modes. (f) For odd multiples of $2\pi$, the superconductor in the
physical region is replaced by a ferromagnet. The domain walls of
the resulting configuration, representing the edges of the original
configurations, give rise to parafermionic bound states \cite{Clarke2013, Lindner2012,Cheng2012}.}
\end{figure*}

Our model allows us to study the physical properties of the FCSC phase.
By considering a Corbino geometry (see Fig. \ref{fig:annulus}), we are able to study $h/2e$ vortices in the system. We show that these vortices host non-abelian parafermion zero-modes, by mapping our configuration to domain walls on the edge of a 2D fractional TI \cite{Lindner2012,Clarke2013,Cheng2012}.

Finally, we examine the Josephson junction depicted in Fig. \hyperref[fig:system]{\ref{fig:system}b}, in which the superconductor stabilizing the FCSC is cut into two concentric annuli.
In the quasi one-dimensional (thin annulus) regime, where coupling between the two edges is allowed, the radial Josephson current is shown to exhibit a $4\pi m$-periodicity as a function of the superconducting phase difference.

\emph{Topological superconductor from an array of coupled wires.---}
We start by reviewing the construction of a $\ensuremath{p_{x}+ip_{y}}$ superconductor, presented in Ref. [\onlinecite{Seroussi2014}].
The starting point is an array of decoupled Rashba
wires. Similar to Refs. \cite{Lutchyn2010,Oreg2010}, each wire is subjected to a Zeeman field
and proximity coupled to an $s$-wave superconductor,
\begin{align}
H  =&\sum_n\int dx\vec{\psi}_{n}^{\dagger}(x)\left[-\frac{(\partial_{x}-iuM\sigma_z)^{2}}{2M}-\mu+B\sigma_{x}\right]\vec{\psi}_{n}(x) \nonumber\\
+&\int dx\left[\Delta\psi_{n\uparrow}^{\dagger}(x)\psi_{n\downarrow}^{\dagger}(x)+{\rm h.c.}\right].\label{eq:basic Hamiltonian}
\end{align}
Here, $\vec{\psi}_{n}=\begin{bmatrix}\psi_{n\uparrow}(x) & \psi_{n\downarrow}(x)\end{bmatrix}^{T}$,
where $\psi_{ns}(x)$ represents the electron annihilation operator
at wire number $n$ with spin $s$ (see Fig. \hyperref[fig:system]{\ref{fig:system}a}). The
matrices $\sigma_{i}$ are Pauli matrices operating on the spin degrees
of freedom. The parameters $u,B,$ and $\Delta$ are the Rashba spin-orbit
coupling (SOC), the Zeeman field, and the induced superconducting pairing potential, respectively.
\begin{figure}
\includegraphics[scale=0.35]{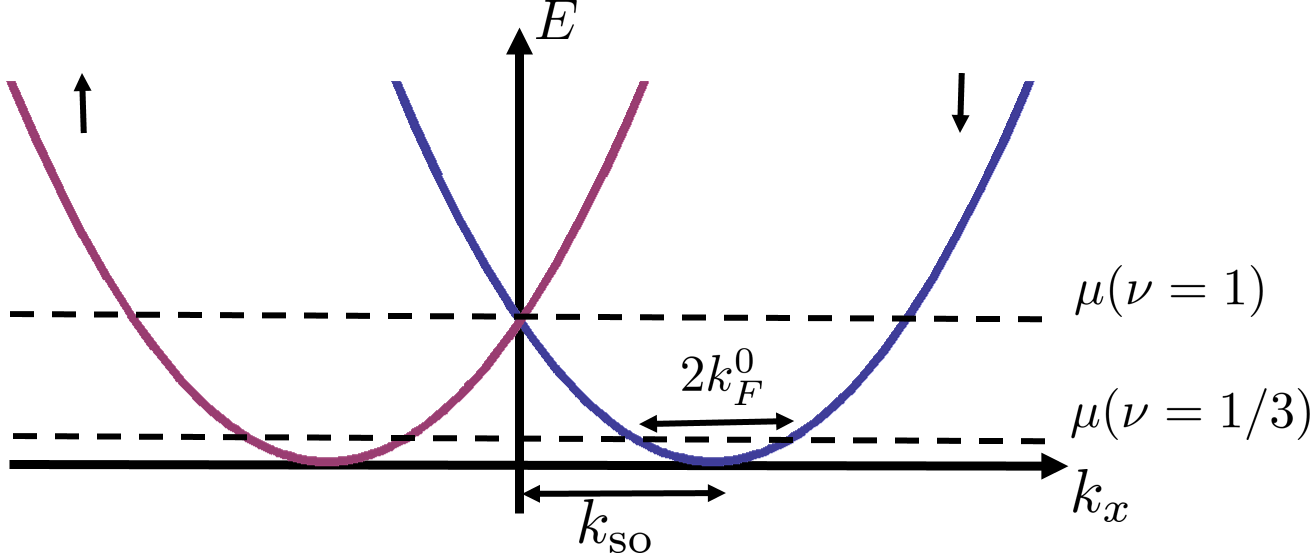}\protect\caption{\label{fig:filling1}The spectrum of an individual wire in the absence
of interactions, Zeeman field, and proximity coupling. The dashed
lines specify the chemical potentials corresponding to the situations
$\nu=1$ and $\nu=1/3$. }
\end{figure}

Figure \ref{fig:filling1} presents the spectrum of a single wire in the absence of superconductivity and a Zeeman field ($B=\Delta=0$). We define the Fermi momentum $k_{F}^{0}=\sqrt{2M\mu}$ in the absence of SOC,
and the shift $k_{\text{so}}=Mu$ of the parabolas due
to SOC. In terms of these, the filling fraction of
the system is defined as
\begin{equation}
\nu=k_{F}^{0}/k_{\text{so}}. \label{eq:filling}
\end{equation}

To
study the low energy physics, we linearize the spectrum near the Fermi level, and decompose the fermionic modes into right- and left-moving
modes:
\begin{equation}
\psi_{ns}(x)=\psi_{nsR}(x)+\psi_{nsL}(x),\label{eq:linearized form}
\end{equation}
with $\psi_{ns\rho}(x)=\psi_{ns\rho}^0(x)e^{ik_{s\rho}x}$. Here, $\psi_{ns\rho}^0$ are the low energy degrees of freedom near the Fermi momenta $k_{s\rho}=\rho k_{F}^{0}-sk_{so}$ (where $\rho=\pm 1$ denotes
a right/left moving mode).

The $p_x+ip_y$ superconductor corresponds to $\nu=1$. Focusing on the low-energy excitations for $\left|\Delta-B\right|\ll B,\Delta$,
the Zeeman term $H_{B}=\int\tx{d}x\left[B\psi_{n\uparrow R}^{\dagger}(x)\psi_{n\downarrow L}(x)+{\rm h.c.}\right]$ and the pairing term $H_{\Delta}=\int\tx{d}x\left\{\Delta\left[\psi_{n\uparrow R}^{\dagger}(x)\psi_{n\downarrow L}^{\dagger}(x)+\psi_{n\uparrow L}^{\dagger}(x)\psi_{n\downarrow R}^{\dagger}(x)\right]+{\rm h.c.}\right\}$ gap out the wires (see Fig. \ref{fig:filling1}).
Indeed, the second term in $H_\Delta$ fully gaps the high-momentum degrees of freedom $\psi_{n\uparrow L}$ and $\psi_{n\downarrow R}$, and the low-momentum degrees of freedom, $\psi_{n\uparrow R}$ and $\psi_{n\downarrow L}$, are coupled by the Zeeman term $H_B$ as well as the first term of $H_\Delta$.
The fully gapped phases for $\Delta>B$ and $\Delta<B$ are topologically
distinct - the first is a 1D trivial phase while the second
is a topological superconductor. For $B=\Delta$, the inter-mode part of the Hamiltonian, given by $H_{\Delta}+H_{B}$,
commutes with the operators

\begin{align}
\alpha_{n}(x) & =\psi_{n\uparrow R}(x)+\psi_{n\uparrow R}^{\dagger}(x),\nonumber \\
\beta_{n}(x) & =-i\left[\psi_{n\downarrow L}(x)-\psi_{n\downarrow L}^{\dagger}(x)\right],\label{eq:Majorana}
\end{align}
so the system is gapless.

Tuning to this transition point,
we introduce an inter-wire term of the form
\begin{equation}
H_{\perp}=it_{\perp}\sum_{n}\int\tx{d}x\alpha_{n}(x)\beta_{n+1}(x).\label{eq:t_perp-1}
\end{equation}
This term arises naturally in the system of Fig. \hyperref[fig:system]{\ref{fig:system}a} due to a combination of inter-wire hopping, superconductivity, and spin-orbit interaction \cite{Seroussi2014}. It gaps out the bulk degrees of freedom,
yet it leaves the Majorana fields $\beta_{1}(x)\text{ and }\alpha_{N}(x)$
decoupled, thus stabilizing a $p_{x}+ip_{y}$ TSC phase. This construction can be considered the anisotropic limit of the system studied by Refs. \cite{Sau2010,Alicea2010}.
While it relies on a fine-tuned relation between parameters, the resulting phase is independent of these assumptions, remaining qualitatively identical as long as the bulk gap remains open.

 \emph{Fractional chiral superconductors.---}
To construct a FCSC phase at filling $\nu=1/m$, we consider
the effects of strong interactions, within a bosonized description
of the 1D degrees of freedom. Using the standard Abelian bosonization
technique, we describe the Hilbert space in terms
of bosonic fields $\phi_{ns\rho}$ via $\psi_{ns\rho}\propto e^{i\left(\phi_{ns\rho}+k_{s\rho}x\right)}$.
The boson fields satisfy the commutation relations
\begin{align}
\left[\phi_{ns\rho}(x),\phi_{n's'\rho'}(x')\right] & =i\pi\rho\delta_{nn'}\delta_{ss'}\delta_{\rho\rho'}\sgn\left(x-x'\right)\label{eq:commutations}\\
 & \hphantom{cccccccccccccccc}+i\pi A_{ns\rho}^{n's'\rho'},\nonumber
\end{align}
 where $A_{ns\rho}^{n's'\rho'}$ takes the values $\pm1$ for any
$\left\{ ns\rho\right\} \neq\left\{ n's'\rho'\right\} $ and is antisymmetric
with respect to the exchange $\left\{ ns\rho\right\} \leftrightarrow\left\{ n's'\rho'\right\} $
(ensuring that the fermionic anti-commutation relations are satisfied).

For $\nu=1/m$, it is useful to define new chiral fermion
operators $\tilde{\psi}_{ns\rho}=e^{i\left(\eta_{ns\rho}+q_{s\rho}x\right)}$, with
\begin{equation}
\eta_{ns\rho}=\frac{m+1}{2}\phi_{ns\rho}-\frac{m-1}{2}\phi_{ns\bar{\rho}}.\label{eq:eta}
\end{equation}
The new fields, $\tilde{\psi}_{ns\rho}$, are fermionic, as evident from the commutation relations
\begin{align}
\left[\eta_{ns\rho}(x),\eta_{n's'\rho'}(x')\right] & =im\pi\rho\delta_{nn'}\delta_{ss'}\delta_{\rho\rho'}\sgn\left(x-x'\right)\label{eq:commutations-1}\\
 & \hphantom{cccccccccccccccc}+i\pi\tilde{A}_{ns\rho}^{n's'\rho'},\nonumber
\end{align}
where $\tilde{A}_{ns\rho}^{n's'\rho'}$ takes odd integer values
for any $\left\{ ns\rho\right\} \neq\left\{ n's'\rho'\right\} $,
and is antisymmetric with respect to the exchange $\left\{ ns\rho\right\} \leftrightarrow\left\{ n's'\rho'\right\} $.
The $\tilde{\psi}_{ns\rho}$ fields carry momenta $q_{s\rho}=\frac{m+1}{2}k_{s\rho}-\frac{m-1}{2}k_{s\bar{\rho}}.$

Note that the momenta $q_{s\rho}$ carried by the new fields $\tilde{\psi}_{ns\rho}$ match those of a $\nu=1$ wire. In particular, the operators
$\tilde{\psi}_{n\uparrow R}$ and $\tilde{\psi}_{n\downarrow L}$ now have
vanishing oscillatory components, allowing us to couple them in various
ways without breaking translational symmetry.

Once the Zeeman
term is dressed by intra-wire $2k_{F}$-interactions, it induces
a term of the form $\tilde{H}_{B}=\int\frac{\tx{d}x}{a^2}\left[\tilde{B}\tilde{\psi}_{n\uparrow R}^{\dagger}(x)\tilde{\psi}_{n\downarrow L}(x)+{\rm h.c.}\right]$ (where $a$ is the short distance cutoff).
We assume that $\tilde{B}$
is large enough such that this term gaps out the small momentum fields
$\tilde{\psi}_{n\uparrow R}\text{ and }\tilde{\psi}_{n\downarrow L}$.
We may also write dressed pairing
terms of the form $\tilde{H}_{\Delta}=\int\frac{\tx{d}x}{a^2}\left\{\tilde{\Delta}\left[\tilde{\psi}_{n\uparrow R}^{\dagger}(x)\tilde{\psi}_{n\downarrow L}^{\dagger}(x)+\tilde{\psi}_{n\uparrow L}^{\dagger}(x)\tilde{\psi}_{n\downarrow R}^{\dagger}(x)\right]+{\rm h.c.}\right\}$. As in the $\nu=1$ case, the second term in $\tilde{H}_\Delta$ involves high-momentum degrees of freedom and does not compete with the Zeeman term, $\tilde{H}_B$. It fully gaps out the fields $\tilde\psi_{n\ua L}$ and $\tilde\psi_{n \da R}$.
The first term in $\tilde{H}_\Delta$ competes
with the Zeeman field $\tilde{H}_B$. For $\tilde{\Delta}=\tilde{B}=\lambda$, one obtains a critical theory, similar to the $\nu=1$ case, which in this case is described by a $\beta^2=4\pi m$ self-dual Sine-Gordon model (see Appendix \ref{appendix:RG}):
\begin{align}
H & =\int dx\left[\left(\partial_{x}\theta\right)^{2}+\left(\partial_{x}\varphi\right)^{2}\right.\nonumber \\
 & \left.\frac{\lambda}{a^{2}}\left\{ \cos\left(\sqrt{4\pi m}\varphi\right)+\cos\left(\sqrt{4\pi m}\theta\right)\right\} \right],\label{eq:SDS}
\end{align}
where we have defined $\varphi =  \frac{\eta_{n\uparrow R}-\eta_{n \downarrow L}}{2\sqrt{\pi m}}$ and $\theta =  \frac{\eta_{n\uparrow R}+\eta_{n \downarrow L}}{2\sqrt{\pi m}}$.

It is, however, a priori unclear whether the critical line $\tilde{B}=\tilde{\Delta}$ is dominated by the $\tilde{B}$ and $\tilde{\Delta}$ terms. In the weak-coupling limit, both $\tilde{B}$ and $\tilde{\Delta}$ flow to zero, giving a trivial Luttinger-liquid fixed point.
However, it turns out that
 when $\tilde{B}$ and $\tilde{\Delta}$ are equal and large enough, a non-trivial multicritical point is encountered.

To show this, we follow the analysis in Ref. [\onlinecite{Boyanovsky1989}], and employ an $\epsilon$-expansion, with $m=2+\epsilon$. In this approach, the scaling dimensions of the $\tilde{B}$ and $\tilde{\Delta}$ terms are small, thus pushing the competition between the first and higher order terms of the RG equations to the region in which the perturbative RG analysis applies. As we show in Appendix \ref{appendix:RG}, the RG equation describing the flow of $\lambda=\tilde{B}=\tilde{\Delta}$ takes the form
 \begin{equation}
\frac{d\lambda}{dl} =  -\epsilon\lambda+\pi^{2}\lambda^3.\label{eq:RG}
\end{equation}
When $\lambda>\sqrt{\epsilon}/\pi$, a flow to large coupling ensues and the low energy theory is no longer capable of describing the model. The point $\lambda = \sqrt{\epsilon}/\pi$ is a multicritical point separating the $\tilde{B}$-dominated phase, the $\tilde{\Delta}$-dominated phase, and the gapless phase. Extrapolating to $\epsilon$ of order unity, we assume that such a critical point persists. For completeness, we study the full phase diagram of the system in Appendix \ref{appendix:RG}.

To uncover the nature of the CFT describing the multicritical point, it is useful to review the physics of classical 2D $\mathbb{Z}_k$ (clock or Potts) models. As discussed in detail in Refs. \cite{Domany1979,Alcaraz1980,Dorey1996}, such models possess self-dual lines. For $k \leq 4$, there is a self-dual critical point that separates the disordered and ordered ($\mathbb{Z}_k$ symmetry broken) phases. For $k>4$, some regions of the self-dual line are contained in gapless phases, while others consist of first order transitions. The different regions of the self-dual line are separated by multicritical points \cite{Domany1979,Dorey1996}. It was argued in Ref. \cite{zamolodchikov1985} that these are described by a $\mathbb{Z}_k$ parafermion CFT. In addition, it is well known that the low energy physics of self-dual $\mathbb{Z}_k$ models is described by $\beta^2=2\pi k$ self-dual Sine-Gordon models \cite{Wiegmann1978, Lecheminant2002}. The above prompts us to identify the finite-coupling multicritical point in our self-dual Sine-Gordon models with those of the $\mathbb{Z}_k$ models. This indicates that our model at the multicritical point is described by a $\mathbb{Z}_{2m}$ parafermion CFT.

 In what follows, we assume that each wire is tuned to the multicritical point, and is therefore described by a $\mathbb{Z}_{2m}$ parafermion theory. At this critical point, the fields
\begin{align}
\tilde{\alpha}_{n}(x) & =e^{i\eta_{n\uparrow R}(x)}+e^{-i\eta_{n\uparrow R}(x)}\nonumber \\
\tilde{\beta}_{n}(x) & =-i\left[e^{i\eta_{n\downarrow L}(x)}-e^{-i\eta_{n\downarrow L}(x)}\right]\label{eq:fractional Majorana}
\end{align}
commute with the inter-mode Hamiltonian $\tilde{H}_{B}+\tilde{H}_{\Delta}$. As we demonstrate in Appendix \ref{appendix:scaling}, the propagators describing
this low energy theory take the form $\left\langle \tilde{\alpha}_{n}(z)\tilde{\alpha}_{n}(z')\right\rangle \propto\left(z-z'\right)^{-m}$ and $\left\langle \tilde{\beta}_{n}(\bar{z})\tilde{\beta}_{n}(\bar{z}')\right\rangle \propto\left(\bar{z}-\bar{z}'\right)^{-m}$, with $z=x+i\tau$. We identify these fields with the $\psi_m$ and $\bar{\psi}_m$ primary fields of the $\mathbb{Z}_{2m}$ theory, which indeed have conformal dimension $m/2$ \cite{zamolodchikov1985}.

Similar to the non-interacting case [see Eq.~\eqref{eq:t_perp-1}], we introduce the inter-wire
term
\begin{equation}
\tilde{H}_{\perp}=i\tilde{t}_{\perp}\sum_{n}\int\frac{\tx{d}x}{a^2}\tilde{\alpha}_{n}(x)\tilde{\beta}_{n+1}(x).\label{eq:t_perp}
\end{equation}
Since the inter-wire Hamiltonian in Eq. (\ref{eq:t_perp}) has the same form as the self-dual intra-wire Hamiltonian and is composed of fields which commute with the intra-wire coupling terms, $\tilde{t}_\perp$ flows according to the RG equation (\ref{eq:RG}). Therefore, if $\tilde{t}_\perp$ is large enough, a flow to large coupling ensues and this Hamiltonian leaves the fields $\tilde{\beta}_{1}$ and $\tilde{\alpha}_{N}$
gapless. These fields represent the local electron operators on the edges. In fact, a full chiral $\mathbb{Z}_{2m}$ parafermion CFT is expected to reside on each edge.
Using the propagator of the $\tilde{\alpha}$ and $\tilde{\beta}$ fields, we find that the electrons' tunneling density of states $N(\omega)$,
associated with the edge, scales with an anomalous exponent:
$N(\omega)\propto\omega^{m-1}$.

Similar to the non-interacting case,
the strict constraints on the various parameters may be lifted as
long as the bulk gap does not close.

\emph{Non-abelian defects and $4\pi m $-periodic Josephson effect.---}
To study the non-Abelian defects residing at the core of vortices, we examine the configuration presented in Fig. \ref{fig:annulus}.
The FCSC is shaped like an annulus, and
the flux threading the annulus is given by an odd multiple of $h/2e$ \footnote{We note that depending on the specific details of the model, the role of even and odd number of vortices may interchange.}. In analogy to the non-interacting
case \cite{Read2000}, we expect to find zero
modes on the two edges of the system. It will prove useful to
cut a thin trench in the annulus. In this case, the flux through the center of the annulus can be chosen to enter in the coupling across the trench, and we can deform our system in
such a way that connects it with the configuration
studied in Refs. \cite{Lindner2012,Clarke2013,Cheng2012} (see Fig. \ref{fig:SFS}).

Given that we are only interested
in finding zero energy modes, we have a large amount of freedom
in deforming the geometry of the problem while preserving its topology. We first fold the edge ``inside-out''
leading to the bone-shaped configuration depicted in Fig. \ref{fig:bone shaped configuration}.
We
may then shrink the outermost regions without introducing inter-edge
coupling. This leads to the simple 1D geometry presented in Fig. \ref{fig:shrinking}, in which the counter-propagating edge modes $\tilde{\alpha}_R$ and $\tilde{\beta}_L$ (defined in the region $0<x<l$) are connected at $x=0,l$.

The zero-mode properties
are solely encoded in the
 low energy edge theory. We therefore have the additional freedom of changing the Hamiltonian governing the gapped degrees of freedom. One must only ensure that the modes $\tilde{\alpha}_{R}$ and $\tilde{\beta}_{L}$, and the other primary fields of the parafermion CFT, remain gapless, while the other microscopic degrees of freedom, such as $\tilde{\alpha}_L$ and $\tilde{\beta}_R$, remain gapped.

An auxiliary model, in which the latter fields are coupled directly, while bulk degrees of freedom are projected out, satisfies this condition.  The corresponding Hamiltonian is given by
\begin{align}
&H_{{\rm auxiliary}} =i\tilde{t}_{\rm{a}}\int_0^l\tx{d}x\tilde{\beta}_{R}(x)\tilde{\alpha}_{L}(x)\nonumber \\
 & =\tilde{t}_{\rm{a}}\int_0^l\tx{d}x\left[\tilde{\psi}_{R}(x)\tilde{\psi}_{L}(x)-\tilde{\psi}_{R}^{\dagger}(x)\tilde{\psi}_{L}(x)+\tx{h.c.}\right],\label{eq:Hunphys}
\end{align}
where we have used $\tilde{\alpha}_\rho=\tilde{\psi}_\rho+\tilde{\psi}_\rho^\dagger$ and $\tilde{\beta}_\rho=\left(\tilde{\psi}_\rho-\tilde{\psi}_\rho^\dagger\right)/i$.  While it gaps $\tilde{\alpha}_L$ and $\tilde{\beta}_R$, this Hamiltonian leaves the fields $\tilde{\alpha}_R,\tilde{\beta}_L$ gapless. Notice that strictly speaking, $\tilde{t}_{\rm{a}}$ must be tuned to a multicritical point for the low-energy theory to remain identical. However, since we will soon break the self duality of the problem, this will not be important.

The auxiliary model simplifies our analysis, as it allows us to formulate the problem in terms of the chiral bosonic modes $\eta_R$ and $\eta_L$, such that $\tilde{\psi}_\rho = e^{i\eta_\rho}$. Mathematically, the Hamiltonian in Eq. (\ref{eq:Hunphys}) describes two edge modes of a $\nu=1/m$
fractional quantum Hall state, coupled by a specific combination of
ferromagnetic and superconducting terms.

 Next, we include the physical coupling between the edge modes across
the trench,
\begin{align}
H_{{\rm trench}} & =i\tilde{t}\left(\varphi\right)\int_0^l\tx{d}x\tilde{\alpha}_{R}(x)\tilde{\beta}_{L}(x)\nonumber \\
 & =\tilde{t}\left(\varphi\right)\int_0^l\tx{d}x\left[\tilde{\psi}_{R}(x)\tilde{\psi}_{L}(x)+\tilde{\psi}_{R}^{\dagger}(x)\tilde{\psi}_{L}(x)+\tx{h.c.}\right],\label{eq:H_phys}
\end{align}
where $\varphi=2\pi\frac{\Phi}{h/2e}$, and $\Phi=nh/2e$ is the flux through the center of the annulus. In the gauge in which the flux only enters into
 $\tilde{t}\left(\varphi\right)$,
we have $\tilde{t}\left(\varphi\right)=\tilde{t}_0\cos\left(\frac{\varphi}{2}\right)$.

Finally, it is necessary
to introduce the correct boundary conditions, such that $\tilde{\alpha}_{R}\rightarrow\tilde{\beta}_{L}$
at $x=0$ and $\tilde{\beta}_{L}\rightarrow\tilde{\alpha}_{R}$ at
$x=l$ (see Fig. \ref{fig:shrinking}).
This can be implemented by extending the model to $-\infty<x<\infty$, and strongly coupling $\alpha_R$ and $\beta_L$ beyond the ends of the physical system, such that they acquire a large gap in these regions.
Within the auxiliary model, formulated in terms of the two $\eta$ fields,
we can do this by introducing strong superconducting terms
for $x<0$ and $x>l$:
\begin{equation}
H_\tx{bc}=\Delta_\tx{bc}\int_{x<0,x>l}\tx{d}x\left[\tilde{\psi}_{R}(x)\tilde{\psi}_{L}(x)+\tx{h.c.}\right].\label{eq:hBC}
\end{equation}
The full auxiliary model is depicted in Fig. \ref{fig:auxiliary model}. Alternatively, note that one can also use a ferromagnetic term. This would not change our conclusions regarding the existence of zero modes.

While the choice of coupling constants $\tilde{t}_a$ and $\Delta_\tx{bc}$
seems arbitrary, unphysical
zero energy degrees of freedom may appear if these are not properly chosen. In order to avoid these,
we first choose the auxiliary coupling constant such that $\sgn\left(\tilde{t}_a\right)=\sgn\left(\Delta_\tx{bc}\right)$
(otherwise the coefficient of the term $i\tilde{\beta}_{R}\tilde{\alpha}_{L}$
changes sign at $x=0,l$).

An additional  consistency condition
is obtained by noting that no physical zero-energy modes should appear
in the absence of flux. This leads to the constraint $\sgn\left(\Delta_\tx{bc}\right)=\sgn\left[\tilde{t}\left(\varphi=0\right)\right]$.

To be consistent with these constraints, the coefficients of our
auxiliary model must satisfy
\begin{equation}
\sgn\left(\tilde{t}_a\right)=\sgn\left(\Delta_\tx{bc}\right)=\sgn\left(\tilde{t}_0\right).\label{eq:condition on auxiliary model}
\end{equation}
Clearly, the gap does not close as long as the various coupling constants do not change sign. In particular, the presence (or absence) of zero-modes
in not affected by varying the various parameters
without altering their signs. We may therefore choose $\tilde{t}_0=\tilde{t}_a\left|\cos\left(\frac{\varphi}{2}\right)\right|^{-1}$,
in which case the Hamiltonian in the region $0<x<l$ takes the form
\begin{align}
H_\tx{trench}+&H_\tx{auxiliary}=\nonumber\\
\tilde{t}_a\int_{0}^l\tx{d}x&\left\{\tilde{\psi}_{R}(x)\tilde{\psi}_{L}(x)\left[1+\sgn\left(\cos\frac{\varphi}{2}\right)\right]\right.
\nonumber\\
&-\left.\tilde{\psi}_{R}^{\dagger}(x)\tilde{\psi}_{L}(x)\left[1-\sgn\left(\cos\frac{\varphi}{2}\right)\right]+\tx{h.c.}\right\}\label{eq:Hamiltonian}
\end{align}

While the trench allows us to continuously vary $\varphi$, we take the discrete values $\varphi=2\pi n$, where $n$ counts the number of $h/2e$ vortices. As Eq. \eqref{eq:Hamiltonian}  implies, if $\varphi$ is an even multiple of $2\pi$, the Hamiltonian reduces to superconducting
terms throughout space, thus preventing the existence of zero energy modes (see Fig. \ref{fig:SFS-1}). On the other hand, if $\varphi$ is an odd multiple of $2\pi$ it is clear that the superconductor in the region $0<x<l$
is replaced by a ferromagnet, and we end up with the S-F-S configuration
studied in Refs. \cite{Lindner2012,Clarke2013,Cheng2012} (see Fig. \ref{fig:SFS}). Notice that if we had chosen to implement the boundary conditions in Eq. \eqref{eq:hBC} using ferromagnetic terms, we would end up with an F-S-F configuration for odd multiples of $2\pi$, and ferromagnetic terms throughout space for even multiple of $2\pi$.

Following
the results of Refs. \cite{Lindner2012,Clarke2013,Cheng2012}, we conclude that the S-F-S and F-S-F configurations found above give rise to parafermion zero-modes $\xi_{i}$, with $i=1,2$, located at the interfaces $x=0,l$. These operators satisfy
\begin{equation}
\xi_{1}\xi_{2}=\xi_{2}\xi_{1}e^{i\frac{\pi}{m}},\label{eq:commutations}
\end{equation}
and $\xi_{i}^{2m}=1$. Eq. \eqref{eq:Hamiltonian} implies that the
spectrum is $4\pi$-periodic as a function of $\varphi$.

We note that the interfaces correspond to the circular edges of the original annulus, so that this line of arguments indicates that $h/2e$ vortices in the FCSC system bind protected parafermionic zero-modes.
The parafermion nature of these excitations provides a natural generalization
to the case where the bulk contains many vortices, each binding a
zero-mode generated by $\xi_{i}$ ($i=1\cdots2M$ is an arbitrary
index labeling the zero-modes). The parafermions generally satisfy
the algebra
\begin{equation}
\xi_{i}\xi_{j}=\xi_{j}\xi_{i}e^{i\frac{\pi}{m}\sgn(i-j)},
\end{equation}
from which one can infer a ground state degeneracy of $\left(2m\right)^{M}$ for $2M$ vortices.

Once our annular system is transformed into the setup presented in
Fig. \ref{fig:SFS} for an odd number of vortices, it is natural to ask whether we can control the
relative phase between the two effective superconducting order-parameters
and measure the Josephson effect, which was previously shown to be
$4\pi m$-periodic \cite{Clarke2013,Cheng2012}.

Recall that the superconductors
shown in Fig. \ref{fig:SFS} were introduced to impose the boundary
conditions, and in particular, do not correspond to the physical superconductor
in proximity to our system. Nevertheless, we argue that such an effective
Josephson junction can be generated by creating a physical Josephson
junction in our system.  This is done by considering a physical weak
link cutting the physical superconductor at a fixed radial coordinate into two concentric annuli
(see Fig. \hyperref[fig:system]{\ref{fig:system}b}). This allows us to control the relative
superconducting phases between the external and the internal parts
of our system by connecting the two edges with an external superconducting
wire, through which flux is inserted.

Since the edges of our annular system correspond to the interfaces in the auxiliary configuration, we effectively control the relative phases between the two auxiliary superconductors shown in Fig. \ref{fig:SFS}   %
\footnote{Notice that the phase differences in the effective and the physical
Josephson junctions play a similar role in changing the wave function
of the low energy electron operators. This indicates that changing
the phase difference in the physical junction results in a phase difference
in the effective junction as well. %
}. Examining the quasi one dimensional regime, in which coupling between the two edge modes of our annular system is induced, the arguments above indicate that the junction shown in Fig. \hyperref[fig:system]{\ref{fig:system}b} gives rise to a $4\pi m$-periodic Josephson effect.

\emph{Conclusion.---} In this work we have demonstrated that a FCSC phase can be stabilized in an array of coupled wires in the presence of Rashba SOC, Zeeman field, superconducting proximity effects, and strong interactions. The resulting phase was found to give rise to a parafermionic edge theory, whose central charge is $c=\frac{2m-1}{m+1}$. In this theory, the electron has an anomalous tunneling density of states of the form $N(\omega)\propto\omega^{m-1}$. We have additionally found that $h/2e$ vortices in such a system bind parafermionic zero modes, with quantum dimension $\sqrt{2m}$.

Clearly, the FCSC phase we construct is topologically ordered. While we leave the detailed study of the bulk excitations to future studies, we anticipate that the topologically distinct bulk excitations will be associated with the various low-energy sectors of the parafermionic edge theory through the bulk-edge correspondence. The low energy sectors are generated by the primary fields of the parafermion CFT, which can be labeled by $\Phi_{[k,\bar{k}]}$ \cite{zamolodchikov1985}, where the integers $k$ and $\bar{k}$ are defined mod $4m$, and $k+\bar{k}$ is even. In terms of these, the parafermion fields are given by $\psi_n=\Phi_{[2n,0]}$ and $\bar{\psi}_n=\Phi_{[0,2n]}$ (with $n=1,\cdots, 2m-1$), and the spin operators $\sigma_n$ are given by $\sigma_n=\Phi_{[n,n]}$.

 However, in identifying the deconfined bulk excitations with the primary fields of the edge CFT, we must exclude the operators which acquire a non-trivial phase as they wind around the electron, which in our case is identified with the field $\psi_{m}$ (or $\bar{\psi}_{m}$). In particular, since the electron acquires a phase of $\pi$ as it winds around a vortex, the corresponding field must be associated with a confined excitation (i.e., the energy of two such excitations diverges as we separate them \footnote{Notice that such confined excitation can be introduced to the system externally. For example, a vortex in a $p_x+ip_y$ superfluid is confined, but it can be induced by applying an external effective magnetic field.} ). More generally, the phase associated with winding $\Phi_{[k,\bar{k}]}$ around the electron $\psi_m$ ($\bar{\psi}_m$) is given by $\gamma_{[k,\bar{k}]}=\pi k$ ($\bar{\gamma}_{[k,\bar{k}]}=\pi \bar{k}$) \cite{zamolodchikov1985}. This shows, for example, that while the parafermion fields $\psi_n$ are all deconfined in the bulk, the spin operators $\sigma_n$, with odd $n$, must be confined.

It has recently been shown that repulsive interactions can stabilize a time reversal invariant topological superconducting (TRITOPS) phase in quantum wires \cite{Gaidamauskas2014,Haim2014,Klinovaja2014, Klinovaja2014c, Danon2015, Haim2016}. It would be interesting to consider an array of such 1D systems as a possible realization of a fractional TRITOPS in 2D.

\smallskip

\begin{acknowledgments}
We are most grateful to Patrick Azaria, Eduardo Fradkin, Michael Levin, Raul Santos, Eran Sela, and Ady Stern for educating us on multiple topics. This work was supported by the Deutsche Forschungsgemeinschaft
 (CRC 183). Y.O. was supported by the Israel
Science Foundation (ISF), the Binational Science
Foundation (BSF), and the European
Research Council under the European Community's Seventh
Framework Program (FP7/2007-2013)/ERC Grant agreement
No. 340210.  E. B. was supported in part by the European
Research Council (ERC) under the Horizon 2020 research and innovation programme
(grant agreement No 639172). E.S. was supported by the Adams Fellowship Program
of the Israel Academy of Sciences and Humanities.

\end{acknowledgments}

\bibliographystyle{apsrev4-1}
%\bibliography{ref}

\begin{thebibliography}{56}%
\makeatletter
\providecommand \@ifxundefined [1]{%
 \@ifx{#1\undefined}
}%
\providecommand \@ifnum [1]{%
 \ifnum #1\expandafter \@firstoftwo
 \else \expandafter \@secondoftwo
 \fi
}%
\providecommand \@ifx [1]{%
 \ifx #1\expandafter \@firstoftwo
 \else \expandafter \@secondoftwo
 \fi
}%
\providecommand \natexlab [1]{#1}%
\providecommand \enquote  [1]{``#1''}%
\providecommand \bibnamefont  [1]{#1}%
\providecommand \bibfnamefont [1]{#1}%
\providecommand \citenamefont [1]{#1}%
\providecommand \href@noop [0]{\@secondoftwo}%
\providecommand \href [0]{\begingroup \@sanitize@url \@href}%
\providecommand \@href[1]{\@@startlink{#1}\@@href}%
\providecommand \@@href[1]{\endgroup#1\@@endlink}%
\providecommand \@sanitize@url [0]{\catcode `\\12\catcode `\$12\catcode
  `\&12\catcode `\#12\catcode `\^12\catcode `\_12\catcode `\%12\relax}%
\providecommand \@@startlink[1]{}%
\providecommand \@@endlink[0]{}%
\providecommand \url  [0]{\begingroup\@sanitize@url \@url }%
\providecommand \@url [1]{\endgroup\@href {#1}{\urlprefix }}%
\providecommand \urlprefix  [0]{URL }%
\providecommand \Eprint [0]{\href }%
\providecommand \doibase [0]{http://dx.doi.org/}%
\providecommand \selectlanguage [0]{\@gobble}%
\providecommand \bibinfo  [0]{\@secondoftwo}%
\providecommand \bibfield  [0]{\@secondoftwo}%
\providecommand \translation [1]{[#1]}%
\providecommand \BibitemOpen [0]{}%
\providecommand \bibitemStop [0]{}%
\providecommand \bibitemNoStop [0]{.\EOS\space}%
\providecommand \EOS [0]{\spacefactor3000\relax}%
\providecommand \BibitemShut  [1]{\csname bibitem#1\endcsname}%
\let\auto@bib@innerbib\@empty
%</preamble>
\bibitem [{\citenamefont {Schnyder}\ \emph {et~al.}(2008)\citenamefont
  {Schnyder}, \citenamefont {Ryu}, \citenamefont {Furusaki},\ and\
  \citenamefont {Ludwig}}]{Schnyder2008}%
  \BibitemOpen
  \bibfield  {author} {\bibinfo {author} {\bibfnamefont {A.}~\bibnamefont
  {Schnyder}}, \bibinfo {author} {\bibfnamefont {S.}~\bibnamefont {Ryu}},
  \bibinfo {author} {\bibfnamefont {A.}~\bibnamefont {Furusaki}}, \ and\
  \bibinfo {author} {\bibfnamefont {A.}~\bibnamefont {Ludwig}},\ }\href
  {\doibase 10.1103/PhysRevB.78.195125} {\bibfield  {journal} {\bibinfo
  {journal} {Physical Review B}\ }\textbf {\bibinfo {volume} {78}},\ \bibinfo
  {pages} {195125} (\bibinfo {year} {2008})}\BibitemShut {NoStop}%
\bibitem [{\citenamefont {Kitaev}(2009)}]{Kitaev2009}%
  \BibitemOpen
  \bibfield  {author} {\bibinfo {author} {\bibfnamefont {A.}~\bibnamefont
  {Kitaev}},\ }\href {http://arxiv.org/abs/arXiv:0901.2686} {\  (\bibinfo
  {year} {2009})},\ \Eprint {http://arxiv.org/abs/0901.2686} {arXiv:0901.2686}
  \BibitemShut {NoStop}%
\bibitem [{\citenamefont {Fu}(2011)}]{Fu2011}%
  \BibitemOpen
  \bibfield  {author} {\bibinfo {author} {\bibfnamefont {L.}~\bibnamefont
  {Fu}},\ }\href {\doibase 10.1103/PhysRevLett.106.106802} {\bibfield
  {journal} {\bibinfo  {journal} {Phys. Rev. Lett.}\ }\textbf {\bibinfo
  {volume} {106}},\ \bibinfo {pages} {106802} (\bibinfo {year}
  {2011})}\BibitemShut {NoStop}%
\bibitem [{\citenamefont {Hsieh}\ \emph {et~al.}(2012)\citenamefont {Hsieh},
  \citenamefont {Lin}, \citenamefont {Liu}, \citenamefont {Duan}, \citenamefont
  {Bansil},\ and\ \citenamefont {Fu}}]{Hsieh2012}%
  \BibitemOpen
  \bibfield  {author} {\bibinfo {author} {\bibfnamefont {T.~H.}\ \bibnamefont
  {Hsieh}}, \bibinfo {author} {\bibfnamefont {H.}~\bibnamefont {Lin}}, \bibinfo
  {author} {\bibfnamefont {J.}~\bibnamefont {Liu}}, \bibinfo {author}
  {\bibfnamefont {W.}~\bibnamefont {Duan}}, \bibinfo {author} {\bibfnamefont
  {A.}~\bibnamefont {Bansil}}, \ and\ \bibinfo {author} {\bibfnamefont
  {L.}~\bibnamefont {Fu}},\ }\href {\doibase 10.1038/ncomms1969} {\bibfield
  {journal} {\bibinfo  {journal} {Nat. Commun.}\ }\textbf {\bibinfo {volume}
  {3}},\ \bibinfo {pages} {982} (\bibinfo {year} {2012})}\BibitemShut {NoStop}%
\bibitem [{\citenamefont {Tanaka}\ \emph {et~al.}(2012)\citenamefont {Tanaka},
  \citenamefont {Ren}, \citenamefont {Sato}, \citenamefont {Nakayama},
  \citenamefont {Souma}, \citenamefont {Takahashi}, \citenamefont {Segawa},\
  and\ \citenamefont {Ando}}]{Tanaka2012}%
  \BibitemOpen
  \bibfield  {author} {\bibinfo {author} {\bibfnamefont {Y.}~\bibnamefont
  {Tanaka}}, \bibinfo {author} {\bibfnamefont {Z.}~\bibnamefont {Ren}},
  \bibinfo {author} {\bibfnamefont {T.}~\bibnamefont {Sato}}, \bibinfo {author}
  {\bibfnamefont {K.}~\bibnamefont {Nakayama}}, \bibinfo {author}
  {\bibfnamefont {S.}~\bibnamefont {Souma}}, \bibinfo {author} {\bibfnamefont
  {T.}~\bibnamefont {Takahashi}}, \bibinfo {author} {\bibfnamefont
  {K.}~\bibnamefont {Segawa}}, \ and\ \bibinfo {author} {\bibfnamefont
  {Y.}~\bibnamefont {Ando}},\ }\href {\doibase 10.1038/nphys2442} {\bibfield
  {journal} {\bibinfo  {journal} {Nat. Phys.}\ }\textbf {\bibinfo {volume}
  {8}},\ \bibinfo {pages} {800} (\bibinfo {year} {2012})}\BibitemShut {NoStop}%
\bibitem [{\citenamefont {Dziawa}\ \emph {et~al.}(2012)\citenamefont {Dziawa},
  \citenamefont {Kowalski}, \citenamefont {Dybko}, \citenamefont {Buczko},
  \citenamefont {Szczerbakow}, \citenamefont {Szot}, \citenamefont
  {{\L}usakowska}, \citenamefont {Balasubramanian}, \citenamefont {Wojek},
  \citenamefont {Berntsen} \emph {et~al.}}]{Dziawa2012}%
  \BibitemOpen
  \bibfield  {author} {\bibinfo {author} {\bibfnamefont {P.}~\bibnamefont
  {Dziawa}}, \bibinfo {author} {\bibfnamefont {B.}~\bibnamefont {Kowalski}},
  \bibinfo {author} {\bibfnamefont {K.}~\bibnamefont {Dybko}}, \bibinfo
  {author} {\bibfnamefont {R.}~\bibnamefont {Buczko}}, \bibinfo {author}
  {\bibfnamefont {A.}~\bibnamefont {Szczerbakow}}, \bibinfo {author}
  {\bibfnamefont {M.}~\bibnamefont {Szot}}, \bibinfo {author} {\bibfnamefont
  {E.}~\bibnamefont {{\L}usakowska}}, \bibinfo {author} {\bibfnamefont
  {T.}~\bibnamefont {Balasubramanian}}, \bibinfo {author} {\bibfnamefont
  {B.~M.}\ \bibnamefont {Wojek}}, \bibinfo {author} {\bibfnamefont
  {M.}~\bibnamefont {Berntsen}},  \emph {et~al.},\ }\href@noop {} {\bibfield
  {journal} {\bibinfo  {journal} {Nat. Mater.}\ }\textbf {\bibinfo {volume}
  {11}},\ \bibinfo {pages} {1023} (\bibinfo {year} {2012})}\BibitemShut
  {NoStop}%
\bibitem [{\citenamefont {Xu}\ \emph {et~al.}(2012)\citenamefont {Xu},
  \citenamefont {Liu}, \citenamefont {Alidoust}, \citenamefont {Neupane},
  \citenamefont {Qian}, \citenamefont {Belopolski}, \citenamefont {Denlinger},
  \citenamefont {Wang}, \citenamefont {Lin}, \citenamefont {Wray} \emph
  {et~al.}}]{Xu2012}%
  \BibitemOpen
  \bibfield  {author} {\bibinfo {author} {\bibfnamefont {S.-Y.}\ \bibnamefont
  {Xu}}, \bibinfo {author} {\bibfnamefont {C.}~\bibnamefont {Liu}}, \bibinfo
  {author} {\bibfnamefont {N.}~\bibnamefont {Alidoust}}, \bibinfo {author}
  {\bibfnamefont {M.}~\bibnamefont {Neupane}}, \bibinfo {author} {\bibfnamefont
  {D.}~\bibnamefont {Qian}}, \bibinfo {author} {\bibfnamefont {I.}~\bibnamefont
  {Belopolski}}, \bibinfo {author} {\bibfnamefont {J.}~\bibnamefont
  {Denlinger}}, \bibinfo {author} {\bibfnamefont {Y.}~\bibnamefont {Wang}},
  \bibinfo {author} {\bibfnamefont {H.}~\bibnamefont {Lin}}, \bibinfo {author}
  {\bibfnamefont {L.}~\bibnamefont {Wray}},  \emph {et~al.},\ }\href@noop {}
  {\bibfield  {journal} {\bibinfo  {journal} {Nat. Commun.}\ }\textbf {\bibinfo
  {volume} {3}},\ \bibinfo {pages} {1192} (\bibinfo {year} {2012})}\BibitemShut
  {NoStop}%
\bibitem [{\citenamefont {Liu}\ \emph {et~al.}(2014)\citenamefont {Liu},
  \citenamefont {Zhang},\ and\ \citenamefont {VanLeeuwen}}]{Liu2014}%
  \BibitemOpen
  \bibfield  {author} {\bibinfo {author} {\bibfnamefont {C.-X.}\ \bibnamefont
  {Liu}}, \bibinfo {author} {\bibfnamefont {R.-X.}\ \bibnamefont {Zhang}}, \
  and\ \bibinfo {author} {\bibfnamefont {B.~K.}\ \bibnamefont {VanLeeuwen}},\
  }\href {\doibase 10.1103/PhysRevB.90.085304} {\bibfield  {journal} {\bibinfo
  {journal} {Phys. Rev. B}\ }\textbf {\bibinfo {volume} {90}},\ \bibinfo
  {pages} {085304} (\bibinfo {year} {2014})}\BibitemShut {NoStop}%
\bibitem [{\citenamefont {Fang}\ and\ \citenamefont {Fu}(2015)}]{Fang2015}%
  \BibitemOpen
  \bibfield  {author} {\bibinfo {author} {\bibfnamefont {C.}~\bibnamefont
  {Fang}}\ and\ \bibinfo {author} {\bibfnamefont {L.}~\bibnamefont {Fu}},\
  }\href {\doibase 10.1103/PhysRevB.91.161105} {\bibfield  {journal} {\bibinfo
  {journal} {Phys. Rev. B}\ }\textbf {\bibinfo {volume} {91}},\ \bibinfo
  {pages} {161105} (\bibinfo {year} {2015})}\BibitemShut {NoStop}%
\bibitem [{\citenamefont {Shiozaki}\ \emph {et~al.}(2015)\citenamefont
  {Shiozaki}, \citenamefont {Sato},\ and\ \citenamefont {Gomi}}]{Shiozaki2015}%
  \BibitemOpen
  \bibfield  {author} {\bibinfo {author} {\bibfnamefont {K.}~\bibnamefont
  {Shiozaki}}, \bibinfo {author} {\bibfnamefont {M.}~\bibnamefont {Sato}}, \
  and\ \bibinfo {author} {\bibfnamefont {K.}~\bibnamefont {Gomi}},\ }\href
  {\doibase 10.1103/PhysRevB.91.155120} {\bibfield  {journal} {\bibinfo
  {journal} {Phys. Rev. B}\ }\textbf {\bibinfo {volume} {91}},\ \bibinfo
  {pages} {155120} (\bibinfo {year} {2015})}\BibitemShut {NoStop}%
\bibitem [{\citenamefont {Wang}\ \emph {et~al.}(2016)\citenamefont {Wang},
  \citenamefont {Alexandradinata}, \citenamefont {Cava},\ and\ \citenamefont
  {Bernevig}}]{Wang2016}%
  \BibitemOpen
  \bibfield  {author} {\bibinfo {author} {\bibfnamefont {Z.}~\bibnamefont
  {Wang}}, \bibinfo {author} {\bibfnamefont {A.}~\bibnamefont
  {Alexandradinata}}, \bibinfo {author} {\bibfnamefont {R.~J.}\ \bibnamefont
  {Cava}}, \ and\ \bibinfo {author} {\bibfnamefont {B.~A.}\ \bibnamefont
  {Bernevig}},\ }\href@noop {} {\bibfield  {journal} {\bibinfo  {journal}
  {Nature}\ }\textbf {\bibinfo {volume} {532}},\ \bibinfo {pages} {189}
  (\bibinfo {year} {2016})}\BibitemShut {NoStop}%
\bibitem [{\citenamefont {Kane}\ \emph {et~al.}(2002)\citenamefont {Kane},
  \citenamefont {Mukhopadhyay},\ and\ \citenamefont {Lubensky}}]{Kane2002}%
  \BibitemOpen
  \bibfield  {author} {\bibinfo {author} {\bibfnamefont {C.}~\bibnamefont
  {Kane}}, \bibinfo {author} {\bibfnamefont {R.}~\bibnamefont {Mukhopadhyay}},
  \ and\ \bibinfo {author} {\bibfnamefont {T.}~\bibnamefont {Lubensky}},\
  }\href {\doibase 10.1103/PhysRevLett.88.036401} {\bibfield  {journal}
  {\bibinfo  {journal} {Phys. Rev. Lett.}\ }\textbf {\bibinfo {volume} {88}},\
  \bibinfo {pages} {036401} (\bibinfo {year} {2002})}\BibitemShut {NoStop}%
\bibitem [{\citenamefont {Teo}\ and\ \citenamefont {Kane}(2014)}]{Teo2014}%
  \BibitemOpen
  \bibfield  {author} {\bibinfo {author} {\bibfnamefont {J.~C.~Y.}\
  \bibnamefont {Teo}}\ and\ \bibinfo {author} {\bibfnamefont {C.~L.}\
  \bibnamefont {Kane}},\ }\href {\doibase 10.1103/PhysRevB.89.085101}
  {\bibfield  {journal} {\bibinfo  {journal} {Phys. Rev. B}\ }\textbf {\bibinfo
  {volume} {89}},\ \bibinfo {pages} {085101} (\bibinfo {year}
  {2014})}\BibitemShut {NoStop}%
\bibitem [{\citenamefont {Klinovaja}\ and\ \citenamefont
  {Loss}(2013)}]{Klinovaja2013c}%
  \BibitemOpen
  \bibfield  {author} {\bibinfo {author} {\bibfnamefont {J.}~\bibnamefont
  {Klinovaja}}\ and\ \bibinfo {author} {\bibfnamefont {D.}~\bibnamefont
  {Loss}},\ }\href {\doibase 10.1103/PhysRevLett.111.196401} {\bibfield
  {journal} {\bibinfo  {journal} {Phys. Rev. Lett.}\ }\textbf {\bibinfo
  {volume} {111}},\ \bibinfo {pages} {196401} (\bibinfo {year} {2013})},\
  \Eprint {http://arxiv.org/abs/1302.6132} {arXiv:1302.6132} \BibitemShut
  {NoStop}%
\bibitem [{\citenamefont {Seroussi}\ \emph {et~al.}(2014)\citenamefont
  {Seroussi}, \citenamefont {Berg},\ and\ \citenamefont {Oreg}}]{Seroussi2014}%
  \BibitemOpen
  \bibfield  {author} {\bibinfo {author} {\bibfnamefont {I.}~\bibnamefont
  {Seroussi}}, \bibinfo {author} {\bibfnamefont {E.}~\bibnamefont {Berg}}, \
  and\ \bibinfo {author} {\bibfnamefont {Y.}~\bibnamefont {Oreg}},\ }\href
  {\doibase 10.1103/PhysRevB.89.104523} {\bibfield  {journal} {\bibinfo
  {journal} {Phys. Rev. B}\ }\textbf {\bibinfo {volume} {89}},\ \bibinfo
  {pages} {104523} (\bibinfo {year} {2014})}\BibitemShut {NoStop}%
\bibitem [{\citenamefont {Neupert}\ \emph {et~al.}(2014)\citenamefont
  {Neupert}, \citenamefont {Chamon}, \citenamefont {Mudry},\ and\ \citenamefont
  {Thomale}}]{Neupert2014}%
  \BibitemOpen
  \bibfield  {author} {\bibinfo {author} {\bibfnamefont {T.}~\bibnamefont
  {Neupert}}, \bibinfo {author} {\bibfnamefont {C.}~\bibnamefont {Chamon}},
  \bibinfo {author} {\bibfnamefont {C.}~\bibnamefont {Mudry}}, \ and\ \bibinfo
  {author} {\bibfnamefont {R.}~\bibnamefont {Thomale}},\ }\href {\doibase
  10.1103/PhysRevB.90.205101} {\bibfield  {journal} {\bibinfo  {journal} {Phys.
  Rev. B}\ }\textbf {\bibinfo {volume} {90}},\ \bibinfo {pages} {205101}
  (\bibinfo {year} {2014})}\BibitemShut {NoStop}%
\bibitem [{\citenamefont {Sagi}\ and\ \citenamefont {Oreg}(2014)}]{Sagi2014}%
  \BibitemOpen
  \bibfield  {author} {\bibinfo {author} {\bibfnamefont {E.}~\bibnamefont
  {Sagi}}\ and\ \bibinfo {author} {\bibfnamefont {Y.}~\bibnamefont {Oreg}},\
  }\href {\doibase 10.1103/PhysRevB.90.201102} {\bibfield  {journal} {\bibinfo
  {journal} {Phys. Rev. B}\ }\textbf {\bibinfo {volume} {90}},\ \bibinfo
  {pages} {201102} (\bibinfo {year} {2014})}\BibitemShut {NoStop}%
\bibitem [{\citenamefont {Klinovaja}\ and\ \citenamefont
  {Tserkovnyak}(2014)}]{Klinovaja2014a}%
  \BibitemOpen
  \bibfield  {author} {\bibinfo {author} {\bibfnamefont {J.}~\bibnamefont
  {Klinovaja}}\ and\ \bibinfo {author} {\bibfnamefont {Y.}~\bibnamefont
  {Tserkovnyak}},\ }\href {\doibase 10.1103/PhysRevB.90.115426} {\bibfield
  {journal} {\bibinfo  {journal} {Phys. Rev. B}\ }\textbf {\bibinfo {volume}
  {90}},\ \bibinfo {pages} {115426} (\bibinfo {year} {2014})}\BibitemShut
  {NoStop}%
\bibitem [{\citenamefont {Meng}\ and\ \citenamefont {Sela}(2014)}]{Meng2014}%
  \BibitemOpen
  \bibfield  {author} {\bibinfo {author} {\bibfnamefont {T.}~\bibnamefont
  {Meng}}\ and\ \bibinfo {author} {\bibfnamefont {E.}~\bibnamefont {Sela}},\
  }\href {\doibase 10.1103/PhysRevB.90.235425} {\bibfield  {journal} {\bibinfo
  {journal} {Phys. Rev. B}\ }\textbf {\bibinfo {volume} {90}},\ \bibinfo
  {pages} {235425} (\bibinfo {year} {2014})}\BibitemShut {NoStop}%
\bibitem [{\citenamefont {Santos}\ \emph {et~al.}(2015)\citenamefont {Santos},
  \citenamefont {Huang}, \citenamefont {Gefen},\ and\ \citenamefont
  {Gutman}}]{Santos2015}%
  \BibitemOpen
  \bibfield  {author} {\bibinfo {author} {\bibfnamefont {R.~A.}\ \bibnamefont
  {Santos}}, \bibinfo {author} {\bibfnamefont {C.-W.}\ \bibnamefont {Huang}},
  \bibinfo {author} {\bibfnamefont {Y.}~\bibnamefont {Gefen}}, \ and\ \bibinfo
  {author} {\bibfnamefont {D.~B.}\ \bibnamefont {Gutman}},\ }\href {\doibase
  10.1103/PhysRevB.91.205141} {\bibfield  {journal} {\bibinfo  {journal} {Phys.
  Rev. B}\ }\textbf {\bibinfo {volume} {91}},\ \bibinfo {pages} {205141}
  (\bibinfo {year} {2015})}\BibitemShut {NoStop}%
\bibitem [{\citenamefont {Sagi}\ \emph {et~al.}(2015)\citenamefont {Sagi},
  \citenamefont {Oreg}, \citenamefont {Stern},\ and\ \citenamefont
  {Halperin}}]{Sagi2015a}%
  \BibitemOpen
  \bibfield  {author} {\bibinfo {author} {\bibfnamefont {E.}~\bibnamefont
  {Sagi}}, \bibinfo {author} {\bibfnamefont {Y.}~\bibnamefont {Oreg}}, \bibinfo
  {author} {\bibfnamefont {A.}~\bibnamefont {Stern}}, \ and\ \bibinfo {author}
  {\bibfnamefont {B.~I.}\ \bibnamefont {Halperin}},\ }\href {\doibase
  10.1103/PhysRevB.91.245144} {\bibfield  {journal} {\bibinfo  {journal} {Phys.
  Rev. B}\ }\textbf {\bibinfo {volume} {91}},\ \bibinfo {pages} {245144}
  (\bibinfo {year} {2015})}\BibitemShut {NoStop}%
\bibitem [{\citenamefont {Gorohovsky}\ \emph {et~al.}(2015)\citenamefont
  {Gorohovsky}, \citenamefont {Pereira},\ and\ \citenamefont
  {Sela}}]{Gorohovsky2015}%
  \BibitemOpen
  \bibfield  {author} {\bibinfo {author} {\bibfnamefont {G.}~\bibnamefont
  {Gorohovsky}}, \bibinfo {author} {\bibfnamefont {R.~G.}\ \bibnamefont
  {Pereira}}, \ and\ \bibinfo {author} {\bibfnamefont {E.}~\bibnamefont
  {Sela}},\ }\href {\doibase 10.1103/PhysRevB.91.245139} {\bibfield  {journal}
  {\bibinfo  {journal} {Phys. Rev. B}\ }\textbf {\bibinfo {volume} {91}},\
  \bibinfo {pages} {245139} (\bibinfo {year} {2015})}\BibitemShut {NoStop}%
\bibitem [{\citenamefont {Meng}\ \emph {et~al.}(2015)\citenamefont {Meng},
  \citenamefont {Neupert}, \citenamefont {Greiter},\ and\ \citenamefont
  {Thomale}}]{Meng2015}%
  \BibitemOpen
  \bibfield  {author} {\bibinfo {author} {\bibfnamefont {T.}~\bibnamefont
  {Meng}}, \bibinfo {author} {\bibfnamefont {T.}~\bibnamefont {Neupert}},
  \bibinfo {author} {\bibfnamefont {M.}~\bibnamefont {Greiter}}, \ and\
  \bibinfo {author} {\bibfnamefont {R.}~\bibnamefont {Thomale}},\ }\href
  {\doibase 10.1103/PhysRevB.91.241106} {\bibfield  {journal} {\bibinfo
  {journal} {Phys. Rev. B}\ }\textbf {\bibinfo {volume} {91}},\ \bibinfo
  {pages} {241106} (\bibinfo {year} {2015})}\BibitemShut {NoStop}%
\bibitem [{\citenamefont {Mross}\ \emph {et~al.}(2015)\citenamefont {Mross},
  \citenamefont {Essin},\ and\ \citenamefont {Alicea}}]{Mross2015}%
  \BibitemOpen
  \bibfield  {author} {\bibinfo {author} {\bibfnamefont {D.~F.}\ \bibnamefont
  {Mross}}, \bibinfo {author} {\bibfnamefont {A.}~\bibnamefont {Essin}}, \ and\
  \bibinfo {author} {\bibfnamefont {J.}~\bibnamefont {Alicea}},\ }\href
  {\doibase 10.1103/PhysRevX.5.011011} {\bibfield  {journal} {\bibinfo
  {journal} {Phys. Rev. X}\ }\textbf {\bibinfo {volume} {5}},\ \bibinfo {pages}
  {011011} (\bibinfo {year} {2015})}\BibitemShut {NoStop}%
\bibitem [{\citenamefont {Meng}(2015)}]{Meng2015a}%
  \BibitemOpen
  \bibfield  {author} {\bibinfo {author} {\bibfnamefont {T.}~\bibnamefont
  {Meng}},\ }\href {\doibase 10.1103/PhysRevB.92.115152} {\bibfield  {journal}
  {\bibinfo  {journal} {Phys. Rev. B}\ }\textbf {\bibinfo {volume} {92}},\
  \bibinfo {pages} {115152} (\bibinfo {year} {2015})}\BibitemShut {NoStop}%
\bibitem [{\citenamefont {Sagi}\ and\ \citenamefont {Oreg}(2015)}]{Sagi2015}%
  \BibitemOpen
  \bibfield  {author} {\bibinfo {author} {\bibfnamefont {E.}~\bibnamefont
  {Sagi}}\ and\ \bibinfo {author} {\bibfnamefont {Y.}~\bibnamefont {Oreg}},\
  }\href {\doibase 10.1103/PhysRevB.92.195137} {\bibfield  {journal} {\bibinfo
  {journal} {Phys. Rev. B}\ }\textbf {\bibinfo {volume} {92}},\ \bibinfo
  {pages} {195137} (\bibinfo {year} {2015})}\BibitemShut {NoStop}%
\bibitem [{\citenamefont {Meng}\ \emph {et~al.}(2016)\citenamefont {Meng},
  \citenamefont {Grushin}, \citenamefont {Shtengel},\ and\ \citenamefont
  {Bardarson}}]{Meng2016}%
  \BibitemOpen
  \bibfield  {author} {\bibinfo {author} {\bibfnamefont {T.}~\bibnamefont
  {Meng}}, \bibinfo {author} {\bibfnamefont {A.~G.}\ \bibnamefont {Grushin}},
  \bibinfo {author} {\bibfnamefont {K.}~\bibnamefont {Shtengel}}, \ and\
  \bibinfo {author} {\bibfnamefont {J.~H.}\ \bibnamefont {Bardarson}},\ }\href
  {http://arxiv.org/abs/1602.08856} {\ ,\ \bibinfo {pages} {14} (\bibinfo
  {year} {2016})},\ \Eprint {http://arxiv.org/abs/1602.08856}
  {arXiv:1602.08856} \BibitemShut {NoStop}%
\bibitem [{\citenamefont {Isobe}\ and\ \citenamefont {Fu}(2015)}]{Isobe2015}%
  \BibitemOpen
  \bibfield  {author} {\bibinfo {author} {\bibfnamefont {H.}~\bibnamefont
  {Isobe}}\ and\ \bibinfo {author} {\bibfnamefont {L.}~\bibnamefont {Fu}},\
  }\href {\doibase 10.1103/PhysRevB.92.081304} {\bibfield  {journal} {\bibinfo
  {journal} {Phys. Rev. B}\ }\textbf {\bibinfo {volume} {92}},\ \bibinfo
  {pages} {081304} (\bibinfo {year} {2015})}\BibitemShut {NoStop}%
\bibitem [{\citenamefont {Sahoo}\ \emph {et~al.}(2015)\citenamefont {Sahoo},
  \citenamefont {Zhang},\ and\ \citenamefont {Teo}}]{Sahoo2015}%
  \BibitemOpen
  \bibfield  {author} {\bibinfo {author} {\bibfnamefont {S.}~\bibnamefont
  {Sahoo}}, \bibinfo {author} {\bibfnamefont {Z.}~\bibnamefont {Zhang}}, \ and\
  \bibinfo {author} {\bibfnamefont {J.~C.~Y.}\ \bibnamefont {Teo}},\ }\href
  {http://arxiv.org/abs/1509.07133} {\ ,\ \bibinfo {pages} {31} (\bibinfo
  {year} {2015})},\ \Eprint {http://arxiv.org/abs/1509.07133}
  {arXiv:1509.07133} \BibitemShut {NoStop}%
\bibitem [{\citenamefont {Iadecola}\ \emph {et~al.}(2016)\citenamefont
  {Iadecola}, \citenamefont {Neupert}, \citenamefont {Chamon},\ and\
  \citenamefont {Mudry}}]{Iadecola2016}%
  \BibitemOpen
  \bibfield  {author} {\bibinfo {author} {\bibfnamefont {T.}~\bibnamefont
  {Iadecola}}, \bibinfo {author} {\bibfnamefont {T.}~\bibnamefont {Neupert}},
  \bibinfo {author} {\bibfnamefont {C.}~\bibnamefont {Chamon}}, \ and\ \bibinfo
  {author} {\bibfnamefont {C.}~\bibnamefont {Mudry}},\ }\href {\doibase
  10.1103/PhysRevB.93.195136} {\bibfield  {journal} {\bibinfo  {journal} {Phys.
  Rev. B}\ }\textbf {\bibinfo {volume} {93}},\ \bibinfo {pages} {195136}
  (\bibinfo {year} {2016})}\BibitemShut {NoStop}%
\bibitem [{\citenamefont {Fuji}\ \emph {et~al.}(2016)\citenamefont {Fuji},
  \citenamefont {He}, \citenamefont {Bhattacharjee},\ and\ \citenamefont
  {Pollmann}}]{Fuji2016}%
  \BibitemOpen
  \bibfield  {author} {\bibinfo {author} {\bibfnamefont {Y.}~\bibnamefont
  {Fuji}}, \bibinfo {author} {\bibfnamefont {Y.-C.}\ \bibnamefont {He}},
  \bibinfo {author} {\bibfnamefont {S.}~\bibnamefont {Bhattacharjee}}, \ and\
  \bibinfo {author} {\bibfnamefont {F.}~\bibnamefont {Pollmann}},\ }\href
  {\doibase 10.1103/PhysRevB.93.195143} {\bibfield  {journal} {\bibinfo
  {journal} {Phys. Rev. B}\ }\textbf {\bibinfo {volume} {93}},\ \bibinfo
  {pages} {195143} (\bibinfo {year} {2016})}\BibitemShut {NoStop}%
\bibitem [{\citenamefont {Huang}\ \emph {et~al.}(2016)\citenamefont {Huang},
  \citenamefont {Chen}, \citenamefont {Gomes}, \citenamefont {Neupert},
  \citenamefont {Chamon},\ and\ \citenamefont {Mudry}}]{Huang2016}%
  \BibitemOpen
  \bibfield  {author} {\bibinfo {author} {\bibfnamefont {P.-H.}\ \bibnamefont
  {Huang}}, \bibinfo {author} {\bibfnamefont {J.-H.}\ \bibnamefont {Chen}},
  \bibinfo {author} {\bibfnamefont {P.~R.~S.}\ \bibnamefont {Gomes}}, \bibinfo
  {author} {\bibfnamefont {T.}~\bibnamefont {Neupert}}, \bibinfo {author}
  {\bibfnamefont {C.}~\bibnamefont {Chamon}}, \ and\ \bibinfo {author}
  {\bibfnamefont {C.}~\bibnamefont {Mudry}},\ }\href {\doibase
  10.1103/PhysRevB.93.205123} {\bibfield  {journal} {\bibinfo  {journal} {Phys.
  Rev. B}\ }\textbf {\bibinfo {volume} {93}},\ \bibinfo {pages} {205123}
  (\bibinfo {year} {2016})}\BibitemShut {NoStop}%
\bibitem [{\citenamefont {Vaezi}(2013)}]{Vaezi2013}%
  \BibitemOpen
  \bibfield  {author} {\bibinfo {author} {\bibfnamefont {A.}~\bibnamefont
  {Vaezi}},\ }\href {\doibase 10.1103/PhysRevB.87.035132} {\bibfield  {journal}
  {\bibinfo  {journal} {Phys. Rev. B}\ }\textbf {\bibinfo {volume} {87}},\
  \bibinfo {pages} {035132} (\bibinfo {year} {2013})}\BibitemShut {NoStop}%
\bibitem [{\citenamefont {Mong}\ \emph {et~al.}(2014)\citenamefont {Mong},
  \citenamefont {Clarke}, \citenamefont {Alicea}, \citenamefont {Lindner},
  \citenamefont {Fendley}, \citenamefont {Nayak}, \citenamefont {Oreg},
  \citenamefont {Stern}, \citenamefont {Berg}, \citenamefont {Shtengel},\ and\
  \citenamefont {Fisher}}]{Mong2014}%
  \BibitemOpen
  \bibfield  {author} {\bibinfo {author} {\bibfnamefont {R.~S.~K.}\
  \bibnamefont {Mong}}, \bibinfo {author} {\bibfnamefont {D.~J.}\ \bibnamefont
  {Clarke}}, \bibinfo {author} {\bibfnamefont {J.}~\bibnamefont {Alicea}},
  \bibinfo {author} {\bibfnamefont {N.~H.}\ \bibnamefont {Lindner}}, \bibinfo
  {author} {\bibfnamefont {P.}~\bibnamefont {Fendley}}, \bibinfo {author}
  {\bibfnamefont {C.}~\bibnamefont {Nayak}}, \bibinfo {author} {\bibfnamefont
  {Y.}~\bibnamefont {Oreg}}, \bibinfo {author} {\bibfnamefont {A.}~\bibnamefont
  {Stern}}, \bibinfo {author} {\bibfnamefont {E.}~\bibnamefont {Berg}},
  \bibinfo {author} {\bibfnamefont {K.}~\bibnamefont {Shtengel}}, \ and\
  \bibinfo {author} {\bibfnamefont {M.~P.~A.}\ \bibnamefont {Fisher}},\ }\href
  {\doibase 10.1103/PhysRevX.4.011036} {\bibfield  {journal} {\bibinfo
  {journal} {Phys. Rev. X}\ }\textbf {\bibinfo {volume} {4}},\ \bibinfo {pages}
  {011036} (\bibinfo {year} {2014})}\BibitemShut {NoStop}%
\bibitem [{\citenamefont {Vaezi}(2014)}]{Vaezi2014}%
  \BibitemOpen
  \bibfield  {author} {\bibinfo {author} {\bibfnamefont {A.}~\bibnamefont
  {Vaezi}},\ }\href {\doibase 10.1103/PhysRevX.4.031009} {\bibfield  {journal}
  {\bibinfo  {journal} {Phys. Rev. X}\ }\textbf {\bibinfo {volume} {4}},\
  \bibinfo {pages} {031009} (\bibinfo {year} {2014})}\BibitemShut {NoStop}%
\bibitem [{\citenamefont {Zamolodchikov}\ and\ \citenamefont
  {Fateev}(1985)}]{zamolodchikov1985}%
  \BibitemOpen
  \bibfield  {author} {\bibinfo {author} {\bibfnamefont {A.~B.}\ \bibnamefont
  {Zamolodchikov}}\ and\ \bibinfo {author} {\bibfnamefont {V.~A.}\ \bibnamefont
  {Fateev}},\ }\href@noop {} {\bibfield  {journal} {\bibinfo  {journal} {Zh.
  Eksp. Teor. Fiz.}\ }\textbf {\bibinfo {volume} {89}},\ \bibinfo {pages} {380}
  (\bibinfo {year} {1985})}\BibitemShut {NoStop}%
\bibitem [{\citenamefont {Lindner}\ \emph {et~al.}(2012)\citenamefont
  {Lindner}, \citenamefont {Berg}, \citenamefont {Refael},\ and\ \citenamefont
  {Stern}}]{Lindner2012}%
  \BibitemOpen
  \bibfield  {author} {\bibinfo {author} {\bibfnamefont {N.~H.}\ \bibnamefont
  {Lindner}}, \bibinfo {author} {\bibfnamefont {E.}~\bibnamefont {Berg}},
  \bibinfo {author} {\bibfnamefont {G.}~\bibnamefont {Refael}}, \ and\ \bibinfo
  {author} {\bibfnamefont {A.}~\bibnamefont {Stern}},\ }\href {\doibase
  10.1103/PhysRevX.2.041002} {\bibfield  {journal} {\bibinfo  {journal} {Phys.
  Rev. X}\ }\textbf {\bibinfo {volume} {2}},\ \bibinfo {pages} {041002}
  (\bibinfo {year} {2012})}\BibitemShut {NoStop}%
\bibitem [{\citenamefont {Clarke}\ \emph {et~al.}(2013)\citenamefont {Clarke},
  \citenamefont {Alicea},\ and\ \citenamefont {Shtengel}}]{Clarke2013}%
  \BibitemOpen
  \bibfield  {author} {\bibinfo {author} {\bibfnamefont {D.~J.}\ \bibnamefont
  {Clarke}}, \bibinfo {author} {\bibfnamefont {J.}~\bibnamefont {Alicea}}, \
  and\ \bibinfo {author} {\bibfnamefont {K.}~\bibnamefont {Shtengel}},\ }\href
  {\doibase 10.1038/ncomms2340} {\bibfield  {journal} {\bibinfo  {journal}
  {Nat. Commun.}\ }\textbf {\bibinfo {volume} {4}},\ \bibinfo {pages} {1348}
  (\bibinfo {year} {2013})}\BibitemShut {NoStop}%
\bibitem [{\citenamefont {Cheng}(2012)}]{Cheng2012}%
  \BibitemOpen
  \bibfield  {author} {\bibinfo {author} {\bibfnamefont {M.}~\bibnamefont
  {Cheng}},\ }\href {\doibase 10.1103/PhysRevB.86.195126} {\bibfield  {journal}
  {\bibinfo  {journal} {Phys. Rev. B}\ }\textbf {\bibinfo {volume} {86}},\
  \bibinfo {pages} {195126} (\bibinfo {year} {2012})}\BibitemShut {NoStop}%
\bibitem [{\citenamefont {Lutchyn}\ \emph {et~al.}(2010)\citenamefont
  {Lutchyn}, \citenamefont {Sau},\ and\ \citenamefont {{Das
  Sarma}}}]{Lutchyn2010}%
  \BibitemOpen
  \bibfield  {author} {\bibinfo {author} {\bibfnamefont {R.~M.}\ \bibnamefont
  {Lutchyn}}, \bibinfo {author} {\bibfnamefont {J.~D.}\ \bibnamefont {Sau}}, \
  and\ \bibinfo {author} {\bibfnamefont {S.}~\bibnamefont {{Das Sarma}}},\
  }\href {\doibase 10.1103/PhysRevLett.105.077001} {\bibfield  {journal}
  {\bibinfo  {journal} {Phys. Rev. Lett.}\ }\textbf {\bibinfo {volume} {105}},\
  \bibinfo {pages} {077001} (\bibinfo {year} {2010})}\BibitemShut {NoStop}%
\bibitem [{\citenamefont {Oreg}\ \emph {et~al.}(2010)\citenamefont {Oreg},
  \citenamefont {Refael},\ and\ \citenamefont {von Oppen}}]{Oreg2010}%
  \BibitemOpen
  \bibfield  {author} {\bibinfo {author} {\bibfnamefont {Y.}~\bibnamefont
  {Oreg}}, \bibinfo {author} {\bibfnamefont {G.}~\bibnamefont {Refael}}, \ and\
  \bibinfo {author} {\bibfnamefont {F.}~\bibnamefont {von Oppen}},\ }\href
  {\doibase 10.1103/PhysRevLett.105.177002} {\bibfield  {journal} {\bibinfo
  {journal} {Phys. Rev. Lett.}\ }\textbf {\bibinfo {volume} {105}},\ \bibinfo
  {pages} {177002} (\bibinfo {year} {2010})}\BibitemShut {NoStop}%
\bibitem [{\citenamefont {Sau}\ \emph {et~al.}(2010)\citenamefont {Sau},
  \citenamefont {Lutchyn}, \citenamefont {Tewari},\ and\ \citenamefont
  {Das~Sarma}}]{Sau2010}%
  \BibitemOpen
  \bibfield  {author} {\bibinfo {author} {\bibfnamefont {J.~D.}\ \bibnamefont
  {Sau}}, \bibinfo {author} {\bibfnamefont {R.~M.}\ \bibnamefont {Lutchyn}},
  \bibinfo {author} {\bibfnamefont {S.}~\bibnamefont {Tewari}}, \ and\ \bibinfo
  {author} {\bibfnamefont {S.}~\bibnamefont {Das~Sarma}},\ }\href {\doibase
  10.1103/PhysRevLett.104.040502} {\bibfield  {journal} {\bibinfo  {journal}
  {Phys. Rev. Lett.}\ }\textbf {\bibinfo {volume} {104}},\ \bibinfo {pages}
  {040502} (\bibinfo {year} {2010})}\BibitemShut {NoStop}%
\bibitem [{\citenamefont {Alicea}(2010)}]{Alicea2010}%
  \BibitemOpen
  \bibfield  {author} {\bibinfo {author} {\bibfnamefont {J.}~\bibnamefont
  {Alicea}},\ }\href {\doibase 10.1103/PhysRevB.81.125318} {\bibfield
  {journal} {\bibinfo  {journal} {Phys. Rev. B}\ }\textbf {\bibinfo {volume}
  {81}},\ \bibinfo {pages} {125318} (\bibinfo {year} {2010})}\BibitemShut
  {NoStop}%
\bibitem [{\citenamefont {Boyanovsky}(1989)}]{Boyanovsky1989}%
  \BibitemOpen
  \bibfield  {author} {\bibinfo {author} {\bibfnamefont {D.}~\bibnamefont
  {Boyanovsky}},\ }\href {http://stacks.iop.org/0305-4470/22/i=13/a=051}
  {\bibfield  {journal} {\bibinfo  {journal} {J. Phys. A: Math. Gen.}\ }\textbf
  {\bibinfo {volume} {22}},\ \bibinfo {pages} {2601} (\bibinfo {year}
  {1989})}\BibitemShut {NoStop}%
\bibitem [{\citenamefont {Domany}\ and\ \citenamefont
  {Riedel}(1979)}]{Domany1979}%
  \BibitemOpen
  \bibfield  {author} {\bibinfo {author} {\bibfnamefont {E.}~\bibnamefont
  {Domany}}\ and\ \bibinfo {author} {\bibfnamefont {E.~K.}\ \bibnamefont
  {Riedel}},\ }\href {\doibase 10.1103/PhysRevB.19.5817} {\bibfield  {journal}
  {\bibinfo  {journal} {Phys. Rev. B}\ }\textbf {\bibinfo {volume} {19}},\
  \bibinfo {pages} {5817} (\bibinfo {year} {1979})}\BibitemShut {NoStop}%
\bibitem [{\citenamefont {Alcaraz}\ and\ \citenamefont
  {Koberle}(1980)}]{Alcaraz1980}%
  \BibitemOpen
  \bibfield  {author} {\bibinfo {author} {\bibfnamefont {F.~C.}\ \bibnamefont
  {Alcaraz}}\ and\ \bibinfo {author} {\bibfnamefont {R.}~\bibnamefont
  {Koberle}},\ }\href {http://stacks.iop.org/0305-4470/13/i=5/a=008} {\bibfield
   {journal} {\bibinfo  {journal} {J. Phys. A}\ }\textbf {\bibinfo {volume}
  {13}},\ \bibinfo {pages} {L153} (\bibinfo {year} {1980})}\BibitemShut
  {NoStop}%
\bibitem [{\citenamefont {Dorey}\ \emph {et~al.}(1996)\citenamefont {Dorey},
  \citenamefont {Tateo},\ and\ \citenamefont {Thompson}}]{Dorey1996}%
  \BibitemOpen
  \bibfield  {author} {\bibinfo {author} {\bibfnamefont {P.}~\bibnamefont
  {Dorey}}, \bibinfo {author} {\bibfnamefont {R.}~\bibnamefont {Tateo}}, \ and\
  \bibinfo {author} {\bibfnamefont {K.~E.}\ \bibnamefont {Thompson}},\ }\href
  {\doibase http://dx.doi.org/10.1016/0550-3213(96)00183-6} {\bibfield
  {journal} {\bibinfo  {journal} {Nucl. Phys. B}\ }\textbf {\bibinfo {volume}
  {470}},\ \bibinfo {pages} {317 } (\bibinfo {year} {1996})}\BibitemShut
  {NoStop}%
\bibitem [{\citenamefont {Wiegmann}(1978)}]{Wiegmann1978}%
  \BibitemOpen
  \bibfield  {author} {\bibinfo {author} {\bibfnamefont {P.~B.}\ \bibnamefont
  {Wiegmann}},\ }\href {http://stacks.iop.org/0022-3719/11/i=8/a=019}
  {\bibfield  {journal} {\bibinfo  {journal} {J. Phys. C: Solid State Phys.}\
  }\textbf {\bibinfo {volume} {11}},\ \bibinfo {pages} {1583} (\bibinfo {year}
  {1978})}\BibitemShut {NoStop}%
\bibitem [{\citenamefont {Lecheminant}\ \emph {et~al.}(2002)\citenamefont
  {Lecheminant}, \citenamefont {Gogolin},\ and\ \citenamefont
  {Nersesyan}}]{Lecheminant2002}%
  \BibitemOpen
  \bibfield  {author} {\bibinfo {author} {\bibfnamefont {P.}~\bibnamefont
  {Lecheminant}}, \bibinfo {author} {\bibfnamefont {A.~O.}\ \bibnamefont
  {Gogolin}}, \ and\ \bibinfo {author} {\bibfnamefont {A.~A.}\ \bibnamefont
  {Nersesyan}},\ }\href {\doibase
  http://dx.doi.org/10.1016/S0550-3213(02)00474-1} {\bibfield  {journal}
  {\bibinfo  {journal} {Nucl. Phys. B}\ }\textbf {\bibinfo {volume} {639}},\
  \bibinfo {pages} {502 } (\bibinfo {year} {2002})}\BibitemShut {NoStop}%
\bibitem [{\citenamefont {Read}\ and\ \citenamefont {Green}(2000)}]{Read2000}%
  \BibitemOpen
  \bibfield  {author} {\bibinfo {author} {\bibfnamefont {N.}~\bibnamefont
  {Read}}\ and\ \bibinfo {author} {\bibfnamefont {D.}~\bibnamefont {Green}},\
  }\href {\doibase 10.1103/PhysRevB.61.10267} {\bibfield  {journal} {\bibinfo
  {journal} {Phys. Rev. B}\ }\textbf {\bibinfo {volume} {61}},\ \bibinfo
  {pages} {10267} (\bibinfo {year} {2000})}\BibitemShut {NoStop}%
\bibitem [{\citenamefont {Gaidamauskas}\ \emph {et~al.}(2014)\citenamefont
  {Gaidamauskas}, \citenamefont {Paaske},\ and\ \citenamefont
  {Flensberg}}]{Gaidamauskas2014}%
  \BibitemOpen
  \bibfield  {author} {\bibinfo {author} {\bibfnamefont {E.}~\bibnamefont
  {Gaidamauskas}}, \bibinfo {author} {\bibfnamefont {J.}~\bibnamefont
  {Paaske}}, \ and\ \bibinfo {author} {\bibfnamefont {K.}~\bibnamefont
  {Flensberg}},\ }\href {\doibase 10.1103/PhysRevLett.112.126402} {\bibfield
  {journal} {\bibinfo  {journal} {Phys. Rev. Lett.}\ }\textbf {\bibinfo
  {volume} {112}},\ \bibinfo {pages} {126402} (\bibinfo {year}
  {2014})}\BibitemShut {NoStop}%
\bibitem [{\citenamefont {Haim}\ \emph {et~al.}(2014)\citenamefont {Haim},
  \citenamefont {Keselman}, \citenamefont {Berg},\ and\ \citenamefont
  {Oreg}}]{Haim2014}%
  \BibitemOpen
  \bibfield  {author} {\bibinfo {author} {\bibfnamefont {A.}~\bibnamefont
  {Haim}}, \bibinfo {author} {\bibfnamefont {A.}~\bibnamefont {Keselman}},
  \bibinfo {author} {\bibfnamefont {E.}~\bibnamefont {Berg}}, \ and\ \bibinfo
  {author} {\bibfnamefont {Y.}~\bibnamefont {Oreg}},\ }\href {\doibase
  10.1103/PhysRevB.89.220504} {\bibfield  {journal} {\bibinfo  {journal} {Phys.
  Rev. B}\ }\textbf {\bibinfo {volume} {89}},\ \bibinfo {pages} {220504}
  (\bibinfo {year} {2014})}\BibitemShut {NoStop}%
\bibitem [{\citenamefont {Klinovaja}\ and\ \citenamefont
  {Loss}(2014{\natexlab{a}})}]{Klinovaja2014}%
  \BibitemOpen
  \bibfield  {author} {\bibinfo {author} {\bibfnamefont {J.}~\bibnamefont
  {Klinovaja}}\ and\ \bibinfo {author} {\bibfnamefont {D.}~\bibnamefont
  {Loss}},\ }\href {\doibase 10.1103/PhysRevLett.112.246403} {\bibfield
  {journal} {\bibinfo  {journal} {Phys. Rev. Lett.}\ }\textbf {\bibinfo
  {volume} {112}},\ \bibinfo {pages} {246403} (\bibinfo {year}
  {2014}{\natexlab{a}})}\BibitemShut {NoStop}%
\bibitem [{\citenamefont {Klinovaja}\ and\ \citenamefont
  {Loss}(2014{\natexlab{b}})}]{Klinovaja2014c}%
  \BibitemOpen
  \bibfield  {author} {\bibinfo {author} {\bibfnamefont {J.}~\bibnamefont
  {Klinovaja}}\ and\ \bibinfo {author} {\bibfnamefont {D.}~\bibnamefont
  {Loss}},\ }\href {\doibase 10.1103/PhysRevB.90.045118} {\bibfield  {journal}
  {\bibinfo  {journal} {Phys. Rev. B}\ }\textbf {\bibinfo {volume} {90}},\
  \bibinfo {pages} {045118} (\bibinfo {year} {2014}{\natexlab{b}})}\BibitemShut
  {NoStop}%
\bibitem [{\citenamefont {Danon}\ and\ \citenamefont
  {Flensberg}(2015)}]{Danon2015}%
  \BibitemOpen
  \bibfield  {author} {\bibinfo {author} {\bibfnamefont {J.}~\bibnamefont
  {Danon}}\ and\ \bibinfo {author} {\bibfnamefont {K.}~\bibnamefont
  {Flensberg}},\ }\href {\doibase 10.1103/PhysRevB.91.165425} {\bibfield
  {journal} {\bibinfo  {journal} {Phys. Rev. B}\ }\textbf {\bibinfo {volume}
  {91}},\ \bibinfo {pages} {165425} (\bibinfo {year} {2015})}\BibitemShut
  {NoStop}%
\bibitem [{\citenamefont {Haim}\ \emph {et~al.}(2016)\citenamefont {Haim},
  \citenamefont {W\"olms}, \citenamefont {Berg}, \citenamefont {Oreg},\ and\
  \citenamefont {Flensberg}}]{Haim2016}%
  \BibitemOpen
  \bibfield  {author} {\bibinfo {author} {\bibfnamefont {A.}~\bibnamefont
  {Haim}}, \bibinfo {author} {\bibfnamefont {K.}~\bibnamefont {W\"olms}},
  \bibinfo {author} {\bibfnamefont {E.}~\bibnamefont {Berg}}, \bibinfo {author}
  {\bibfnamefont {Y.}~\bibnamefont {Oreg}}, \ and\ \bibinfo {author}
  {\bibfnamefont {K.}~\bibnamefont {Flensberg}},\ }\href {\doibase
  10.1103/PhysRevB.94.115124} {\bibfield  {journal} {\bibinfo  {journal} {Phys.
  Rev. B}\ }\textbf {\bibinfo {volume} {94}},\ \bibinfo {pages} {115124}
  (\bibinfo {year} {2016})}\BibitemShut {NoStop}%
\end{thebibliography}
%

\begin{appendix}

\section{The renormalization group equations of the self-dual Sine-Gordon
model}\label{appendix:RG}

\subsection{The $\epsilon$ expansion }

The Hamiltonians describing either the inter- and intra- wire coupling
terms, discussed in the main text, can be written as
\begin{align*}
H & =\int dx\left[\frac{1}{2\pi m}\left(\partial_{x}\eta_{1}\right)^{2}+\frac{1}{2\pi m}\left(\partial_{x}\eta_{2}\right)^{2}\right.\\
 & \left.+\frac{\tilde{B}}{a^{2}}\cos\left(\eta_{1}-\eta_{2}\right)+\frac{\tilde{\Delta}}{a^{2}}\cos\left(\eta_{1}+\eta_{2}\right)\right],
\end{align*}
where $\eta_{1}$ ($\eta_{2}$) is a right (left) moving mode satisfying
\begin{eqnarray*}
\left[\eta_{j}(x),\eta_{j}(x')\right] & = & m\pi i(-1)^{j}\text{sign}\left(x-x'\right),
\end{eqnarray*}
and the units were chosen such that the Fermi-velocity is $v=1$.
Specifically, in the main text, the coefficients were tuned to the
self dual line, such that $\tilde{B}=\tilde{\Delta}$.

To write the Hamiltonian in a more convenient form, we define
\begin{eqnarray*}
\varphi & = & \frac{\eta_{1}-\eta_{2}}{2\sqrt{\pi m}}\\
\theta & = & \frac{\eta_{1}+\eta_{2}}{2\sqrt{\pi m}}.
\end{eqnarray*}
The commutation relations of these fields are given by
\[
\left[\varphi(x),\theta(x')\right]=i\Theta(x-x').
\]
In terms of these, the Hamiltonian takes the form
\begin{align*}
H & =\int dx\left[\left(\partial_{x}\theta\right)^{2}+\left(\partial_{x}\varphi\right)^{2}\right.\\
 & \left.+\frac{\tilde{B}}{a^{2}}\cos\left(2\sqrt{\pi m}\varphi\right)+\frac{\tilde{\Delta}}{a^{2}}\cos\left(2\sqrt{\pi m}\theta\right)\right]
\end{align*}

We next wish to isolate the explicit cutoff dependence by writing
the cosines in terms of their normal-ordered versions:
\[
\cos\left(2\sqrt{\pi m}\varphi\right)=:\cos\left(2\sqrt{\pi m}\varphi\right):e^{-2m\pi\left\langle \varphi^{2}\right\rangle }
\]
\[
\cos\left(2\sqrt{\pi m}\theta\right)=:\cos\left(2\sqrt{\pi m}\theta\right):e^{-2m\pi\left\langle \theta^{2}\right\rangle },
\]
where the averages are taken with respect to the quadratic part of
the Hamiltonian.

Denoting the small distance cutoff by $a$ and the large distance
cutoff by $L,$ we find that
\[
e^{-2m\pi\left\langle \varphi^{2}\right\rangle }=e^{-2m\pi\left\langle \theta^{2}\right\rangle }=\left(\frac{a}{L}\right)^{m}.
\]
Therefore, the Hamiltonian can be written in the form
\begin{eqnarray*}
H & = & \int dx\left[\left(\partial_{x}\theta\right)^{2}+\left(\partial_{x}\varphi\right)^{2}\right.\\
 &  & +\frac{\tilde{B}}{L^{2}}\left(\frac{a}{L}\right)^{m-2}:\cos\left(2\sqrt{\pi m}\varphi\right):\\
 &  & \left.+\frac{\tilde{\Delta}}{L^{2}}\left(\frac{a}{L}\right)^{m-2}:\cos\left(2\sqrt{\pi m}\theta\right):\right].
\end{eqnarray*}

In performing the renormalization group, we quantify the dependence
of the coefficients on $l=\log\frac{L}{a}$.

At tree level, we simply use the explicit cutoff dependence of the
coefficients to calculate $\frac{d\bar{B}}{dl},\frac{d\bar{\Delta}}{dl}$,
with $\bar{B}=\tilde{B}\left(\frac{a}{L}\right)^{m-2},\bar{\Delta}=\tilde{\Delta}\left(\frac{a}{L}\right)^{m-2}$.
The resulting tree level RG equations are given by
\begin{eqnarray*}
\frac{d\bar{B}}{dl} & = & \left(2-m\right)\bar{B}\\
\frac{d\bar{\Delta}}{dl} & = & (2-m)\bar{\Delta},
\end{eqnarray*}

giving a flow to weak coupling for $m>2$. This is true in particular
at the self-dual point $\tilde{B}=\tilde{\Delta}=\tilde{\lambda}$
we wish to focus on.

Clearly, if a flow to large coupling is to occur for large $\lambda$,
it must result from higher order terms. The next possible non-vanishing
correction to the RG equation of $\lambda$ is of order $\lambda^{3}$,
as the second order correction vanishes at the self-dual line. Indeed,
if the $\lambda^{3}$ term has a positive coefficient, it induces
a flow to large couplings for large $\lambda$. However, for such
a term to overcome the negative contribution of the first order term,
$\lambda$ should be of order 1 (assuming the coefficients are of
order 1 and recalling that in our model $m=3,5,$ etc.). Such a situation
steps beyond the range of validity of the perturbative RG analysis.

To overcome the above technical difficulty, we study the situation
in which $m$ is slightly above its marginal value: $m=2+\epsilon$,
with $\epsilon\ll1$. In this case, the negative first order contribution
is proportional to $\epsilon$:
\begin{eqnarray*}
\frac{d\bar{B}}{dl} & = & -\epsilon B\\
\frac{d\bar{\Delta}}{dl} & = & -\epsilon\Delta,
\end{eqnarray*}
and therefore, if the third order contribution is positive, the critical
point between the two regimes is controlled by the small parameter
$\epsilon$ and is therefore expected to be within the range of validity
of the perturbative RG analysis. Indeed, we will find such a critical
point satisfying $\lambda_{c}\propto\sqrt{\epsilon}.$

The critical point found for small $\epsilon$ indicates that a similar
critical point exists for $\epsilon=1$ as well, above which the coupling
constants flows to large coupling (assuming no additional critical
points are induced as $\epsilon$ is increased from $0<\epsilon\ll1$
to $\epsilon=1$).

We next turn to explicitly derive the form of the third order RG equations.
The analysis presented below closely parallels the analysis presented
in Ref. \cite{Boyanovsky1989}.

\subsection{Third order RG equations and the phase diagram}

To calculate the higher orders of the renormalization group equations,
we generally write the partition function as
\[
Z=\int D\varphi e^{-S_{0}-S_{1}},
\]
with
\begin{eqnarray*}
S_{0} & =\frac{1}{2} & \int d\tau dx\left[\left(\partial_{\tau}\varphi\right)^{2}+\left(\partial_{x}\varphi\right)^{2}\right].
\end{eqnarray*}

\begin{eqnarray*}
S_{1} & = & \int\frac{d\tau dx}{L^{2}}\left[\bar{B}:\cos\left(\sqrt{8\pi\left(1+\delta_{B}\right)}\varphi\right):\right.\\
 &  & \left.+\bar{\Delta}:\cos\left(\sqrt{8\pi\left(1+\delta_{\Delta}\right)}\theta\right):\right],
\end{eqnarray*}
where $\theta$ is related to $\varphi$ according to the condition:
\begin{equation}
i\partial_{\mu}\varphi=\epsilon^{\mu\nu}\partial_{\nu}\theta.\label{eq:constraint on theta}
\end{equation}
In addition, we have defined $\bar{B}=\tilde{B}\left(\frac{a}{L}\right)^{2\delta_{B}},\bar{\Delta}=\tilde{\Delta}\left(\frac{a}{L}\right)^{2\delta_{\Delta}}$.

In order to follow the strategy outlined above, we assume $\delta_{B}$
and $\delta_{\Delta}$ are small. As these parameters flow as well,
we will write the third order RG equations for $\bar{B},\bar{\Delta},\delta_{B},\delta_{\Delta}$\@.
Once these are derived, we will focus on the self-dual line defined
according to $\bar{B}=\bar{\Delta}=\bar{\lambda}$ and $\delta_{B}=\delta_{\Delta}=\frac{\epsilon}{2}$.

To derive the RG equations, we expand the partition function in orders
of $S_{1}$:
\begin{equation}
Z=\int D\varphi D\theta e^{-S_{0}}\left(1-S_{I}+\frac{1}{2}S_{I}^{2}-\frac{1}{6}S_{I}^{3}+\cdots\right).\label{eq:perturbation}
\end{equation}
Next, we will use the Operator Product Expansions (OPEs) of the resulting
high orders to get corrections to the original action.

Taking only the first order into account, we get the contribution
\begin{eqnarray*}
\frac{d\bar{B}}{dl} & = & -2\delta_{B}\bar{B}\\
\frac{d\bar{\Delta}}{dl} & = & -2\delta_{\Delta}\bar{\Delta}.
\end{eqnarray*}
To second order, terms of the form $\bar{B}^{2}$ and $\bar{\Delta}^{2}$
renormalize the Kinetic term. Notice that we discard the non-singular
term of order $\bar{B}\bar{\Delta}$, which results in an irrelevant
term.

We first calculate the $\bar{B}^{2}$ component of $S_{2}=-\frac{1}{2}S_{I}^{2}$:
\begin{widetext}
\begin{eqnarray}
-\frac{\bar{B}^{2}}{2}\int d^{2}x_{1}d^{2}x_{2}L^{-4}:\cos\left(\sqrt{8\pi\left(1+\delta_{B}\right)}\varphi(\vec{x}_{1})\right)::\cos\left(\sqrt{8\pi\left(1+\delta_{B}\right)}\varphi(\vec{x}_{2})\right): & .\label{eq:B^2}
\end{eqnarray}
In order to calculate the OPE \cite{Boyanovsky1989}, we use the
definition of normal ordering to write
\begin{align*}
 & :\cos\left(\sqrt{8\pi\left(1+\delta_{B}\right)}\varphi(\vec{x}_{1})\right)::\cos\left(\sqrt{8\pi\left(1+\delta_{B}\right)}\varphi(\vec{x}_{2})\right):\\
= & e^{8\pi(1+\delta_{B})\left\langle \varphi^{2}\right\rangle }\cos\left(\sqrt{8\pi\left(1+\delta_{B}\right)}\varphi(\vec{x}_{1})\right)\cos\left(\sqrt{8\pi\left(1+\delta_{B}\right)}\varphi(\vec{x}_{2})\right).
\end{align*}
This can be rewritten as
\[
\frac{1}{2}e^{8\pi(1+\delta_{B})\left\langle \varphi^{2}\right\rangle }\left[\cos\left(\sqrt{8\pi\left(1+\delta_{B}\right)}\left\{ \varphi(\vec{x}_{1})+\varphi(\vec{x}_{2})\right\} \right)+\cos\left(\sqrt{8\pi\left(1+\delta_{B}\right)}\left\{ \varphi(\vec{x}_{1})-\varphi(\vec{x}_{2})\right\} \right)\right].
\]
\end{widetext}
Writing the two cosines in terms of their normal ordered versions,
we get
\begin{align*}
 & \frac{1}{2}\left[:\cos\left(\sqrt{8\pi\left(1+\delta_{B}\right)}\left\{ \varphi(\vec{x}_{1})+\varphi(\vec{x}_{2})\right\} \right):\frac{1}{c(\vec{x}_{1}-\vec{x}_{2})}\right.\\
 & \left.+:\cos\left(\sqrt{8\pi\left(1+\delta_{B}\right)}\left\{ \varphi(\vec{x}_{1})-\varphi(\vec{x}_{2})\right\} \right):c(\vec{x}_{1}-\vec{x}_{2})\right]
\end{align*}
with
\[
c(\vec{x}_{1}-\vec{x}_{2})=e^{8\pi(1+\delta_{B})\left\langle \varphi(\vec{x}_{1})\varphi(\vec{x}_{2})\right\rangle }=\left(\frac{\left|\vec{x}_{1}-\vec{x}_{2}\right|}{L}\right)^{-4(1+\delta_{B})}.
\]
The first term results in irrelevant terms, and is therefore ignored.
The second term is dominated by the region in which the two points
$\vec{x}_{1}\text{ and }\vec{x}_{2}$ are close to each other. This
allows us to approximate
\begin{align*}
 & \cos\left(\sqrt{8\pi\left(1+\delta_{B}\right)}\left\{ \varphi(\vec{x}_{1})-\varphi(\vec{x}_{2})\right\} \right)\\
 & \approx1-4\pi(1+\delta_{B})\left(\varphi(\vec{x}_{1})-\varphi(\vec{x}_{2})\right)^{2}\\
 & =1-4\pi(1+\delta_{B})\left(\left(\vec{x}_{1}-\vec{x}_{2}\right)\cdot\nabla\varphi(\vec{x})\right)^{2},
\end{align*}

where $\vec{x}$ is the center of mass coordinate. Plugging this
back into Eq. (\ref{eq:B^2}), and performing the integral over the
relative coordinate $\vec{x}_{1}-\vec{x}_{2}$, we get a correction
of the form
\[
2\pi^{2}\log\frac{L}{a}\bar{B}^{2}\left(1+\delta_{B}\right)\left[\frac{1}{2}\int\left(\left(\partial_{x}\varphi\right)^{2}+\left(\partial_{\tau}\varphi\right)^{2}\right)d^{2}x\right].
\]

Similarly, the $\bar{\Delta}^{2}$ terms gives us (up to a constant)
\begin{eqnarray}
 &  & :\cos\left(\sqrt{8\pi\left(1+\delta_{\Delta}\right)}\theta(\vec{x}_{1})\right)::\cos\left(\sqrt{8\pi\left(1+\delta_{\Delta}\right)}\theta(\vec{x}_{2})\right):\nonumber \\
 &  & =-\frac{\pi\left(1+\delta_{\Delta}\right)\left|\vec{x}_{1}-\vec{x}_{2}\right|^{2}}{\left(\frac{\left|\vec{x}_{1}-\vec{x}_{2}\right|}{L}\right)^{4(1+\delta_{\Delta})}}\left[\left(\partial_{x}\theta(\vec{x})\right)^{2}+\left(\partial_{\tau}\theta(\vec{x})\right)^{2}\right].\label{eq:B^2-1}
\end{eqnarray}
Using Eq. (\ref{eq:constraint on theta}) and integrating over the
relative coordinate, we obtain the correction
\[
-2\pi^{2}\log\frac{L}{a}\bar{\Delta}^{2}\left(1+\delta_{\Delta}\right)\left[\frac{1}{2}\int\left(\left(\partial_{x}\varphi\right)^{2}+\left(\partial_{\tau}\varphi\right)^{2}\right)d^{2}x\right].
\]
Taken together, the second order corrections are given by
\begin{align*}
S_{2} & =2\pi^{2}\log\frac{L}{a}\left\{ \bar{B}^{2}\left(1+\delta_{B}\right)-\bar{\Delta}^{2}\left(1+\delta_{\Delta}\right)\right\} \\
 & \times\left[\frac{1}{2}\int\left(\left(\partial_{x}\varphi\right)^{2}+\left(\partial_{\tau}\varphi\right)^{2}\right)d^{2}x\right].
\end{align*}

The second order term clearly renormalizes the kinetic part of the
action.

The third order contributions are given by a combination of $S_{3}^{a}=\frac{1}{6}S_{1}^{3}$
and $S_{3}^{b}=S_{1}S_{2}$ (the latter contribution arises when one
re-exponentiates the partition function in Eq. \ref{eq:perturbation}).

We first look at the $S_{3}^{a}$-term. The corrections of the type
$\bar{B}^{3}$ and $\bar{B}\bar{\Delta}^{2}$ renormalize the $B$-term
while terms of the form $\bar{\Delta}^{3}$ and $\bar{\Delta}\bar{B}^{2}$
renormalize the $\Delta$-term. Let us first look at the $\bar{B}^{3}$
order term:
\[
\frac{1}{6}\bar{B}^{3}\int\Pi_{n=1}^{3}\left[\frac{d^{2}x_{n}}{L^{2}}:\cos\left(\sqrt{8\pi\left(1+\delta_{B}\right)}\varphi(\vec{x}_{n})\right):\right].
\]
Using the OPE and dropping the non-singular term, we get the correction
\cite{Boyanovsky1989}

\begin{eqnarray*}
 &  & \frac{\bar{B}^{3}}{8L^{6}}\int d^{2}x_{1}d^{2}x_{2}d^{2}x_{3} \left(\frac{L^{2}\left|\vec{x}_{1}-\vec{x}_{2}\right|^{2}}{\left|\vec{x}_{1}-\vec{x}_{3}\right|^{2}\left|\vec{x}_{2}-\vec{x}_{3}\right|^{2}}\right)^{m}\\
 &  & \times :\cos\left(\sqrt{8\pi\left(1+\delta_{B}\right)}\left(\varphi(\vec{x}_{1})+\varphi(\vec{x}_{2})-\varphi(\vec{x}_{3})\right)\right):.
\end{eqnarray*}

Clearly, the most dominant contributions arise when two coordinates
approach each other: either $\vec{x}_{1}\rightarrow\vec{x}_{3}$,
or $\vec{x}_{2}\rightarrow\vec{x}_{3}$. However, these divergences
are disconnected and are therefore discarded. Thus, the only singular
contribution arises when the three different coordinates approach
each other. Taking this into account, we approximate
\begin{align*}
 & :\cos\left(\sqrt{8\pi\left(1+\delta_{B}\right)}\left(\varphi(\vec{x}_{1})+\varphi(\vec{x}_{2})-\varphi(\vec{x_{3})}\right)\right):\\
 & \approx:\cos\left(\sqrt{8\pi\left(1+\delta_{B}\right)}\varphi(\vec{x})\right):,
\end{align*}
where $\vec{x}$ is the center of mass coordinate, and therefore
\begin{align*}
 & \frac{1}{8L^{6}}\int d^{2}x_{1}d^{2}x_{2}d^{2}x_{3}\left(\frac{L^{2}\left|\vec{x}_{1}-\vec{x}_{2}\right|^{2}}{\left|\vec{x}_{1}-\vec{x}_{3}\right|^{2}\left|\vec{x}_{2}-\vec{x}_{3}\right|^{2}}\right)^{2}\\
 & \times:\cos\left(\sqrt{8\pi\left(1+\delta_{B}\right)}\varphi(\vec{x})\right):
\end{align*}
Integrating over the relative coordinates, and subtracting disconnected
terms, we find the correction \cite{Boyanovsky1989}
\[
\frac{2\pi^{2}\bar{B}^{3}}{L^{2}}\left[\log\left(\frac{L}{a}\right)\right]^{2}\int d^{2}x:\cos\left(\sqrt{8\pi\left(1+\delta_{B}\right)}\varphi(\vec{x})\right):.
\]

Similar considerations show that the $\bar{B}\bar{\Delta}^{2}$ term
provides a contribution of the form
\begin{align*}
 & \frac{\pi^{2}\bar{B}\bar{\Delta}^{2}}{L^{2}}\left[\log\left(\frac{L}{a}\right)-2\left[\log\left(\frac{L}{a}\right)\right]^{2}\right]\\
 & \times\int d^{2}x:\cos\left(\sqrt{8\pi\left(1+\delta_{B}\right)}\varphi(\vec{x})\right):.
\end{align*}
The remaining contribution is of the form $S_{3}^{b}$. The contribution
to the $B$-term is given by
\begin{widetext}
\begin{eqnarray*}
 &  & 2\pi^{2}\log\frac{L}{a}\bar{B}\left\{ \bar{B}^{2}\left(1+\delta_{B}\right)-\bar{\Delta}^{2}\left(1+\delta_{\Delta}\right)\right\} \\
 &  & \times\left[\frac{1}{2}\int d^{2}x_{1}d^{2}x_{2}\mu^{2}\left(\left(\partial_{x}\varphi(\vec{x}_{1})\right)^{2}+\left(\partial_{\tau}\varphi(\vec{x}_{1})\right)^{2}\right):\cos\left(\sqrt{8\pi\left(1+\delta_{B}\right)}\varphi(\vec{x}_{2})\right):\right].
\end{eqnarray*}
Keeping only the third orders in the various scaling parameters, and
using the corresponding OPE, we get
\[
-2\pi^{2}\left(\log\frac{L}{a}\right)^{2}\bar{B}\left\{ \bar{B}^{2}-\bar{\Delta}^{2}\right\} \int d^{2}x:\cos\left(\sqrt{8\pi\left(1+\delta_{B}\right)}\left(\varphi(\vec{x})\right)\right):.
\]
Analogous corrections can be derived for the $\Delta$-field.

Summing the above corrections, we find that the renormalized action
is given by
\begin{eqnarray*}
S & = & S_{I}+2\pi^{2}\log\frac{L}{a}\left\{ \bar{B}^{2}\left(1+\delta_{B}\right)-\bar{\Delta}^{2}\left(1+\delta_{\Delta}\right)\right\} \left[\frac{1}{2}\int\left(\left(\partial_{x}\varphi\right)^{2}+\left(\partial_{\tau}\varphi\right)^{2}\right)d^{2}x\right]\\
 &  & +\frac{\pi^{2}\bar{B}\bar{\Delta}^{2}}{L^{2}}\log\left(\frac{L}{a}\right)\int d^{2}x:\cos\left(\sqrt{8\pi\left(1+\delta_{B}\right)}\varphi(\vec{x})\right):\\
 &  & +\frac{\pi^{2}\bar{\Delta}\bar{B}^{2}}{L^{2}}\log\left(\frac{L}{a}\right)\int d^{2}x:\cos\left(\sqrt{8\pi\left(1+\delta_{\Delta}\right)}\theta(\vec{x})\right):.
\end{eqnarray*}
\end{widetext}
Based on the above, the RG equations for the coefficients of the cosine
terms are given by
\begin{eqnarray*}
\frac{d\bar{B}}{dl} & = & -2\delta_{B}\bar{B}+\pi^{2}\bar{B}\bar{\Delta}^{2}\\
\frac{d\bar{\Delta}}{dl} & = & -2\delta_{\Delta}\bar{\Delta}+\pi^{2}\bar{\Delta}\bar{B}^{2}.
\end{eqnarray*}
The correction to the Kinetic term can be eliminated by rescaling
the $\varphi$ and $\theta$-fields. This induces a change in the
variables $\delta_{B}$, $\delta_{\Delta}$, leading to the RG equations:
\begin{eqnarray*}
\frac{d\delta_{B}}{dl} & = & -2\pi^{2}(1+\delta_{B})\left[\bar{B}^{2}\left(1+\delta_{B}\right)-\bar{\Delta}^{2}\left(1+\delta_{\Delta}\right)\right]\\
\frac{d\delta_{\Delta}}{dl} & = & -2\pi^{2}(1+\delta_{\Delta})\left[\bar{\Delta}^{2}\left(1+\delta_{\Delta}\right)-\bar{B}^{2}\left(1+\delta_{B}\right)\right].
\end{eqnarray*}
Rescaling $\bar{B}'=\pi^{2}\bar{B},\bar{\Delta}'=\pi^{2}\bar{\Delta},$
we finally write the RG equations
\begin{eqnarray*}
\frac{d\bar{B}'}{dl} & = & -2\delta_{B}\bar{B}'+\bar{B}'\bar{\Delta}'^{2}\\
\frac{d\bar{\Delta}}{dl} & = & -2\delta_{\Delta}\bar{\Delta}'+\bar{\Delta}'\bar{B}'^{2}
\end{eqnarray*}
\begin{eqnarray}
\frac{d\delta_{B}}{dl} & = & -2(1+\delta_{B})\left[\bar{B}'^{2}(1+\delta_{B})-\bar{\Delta}'^{2}(1+\delta_{\Delta})\right]\nonumber \\
\frac{d\delta_{\Delta}}{dl} & = & -2(1+\delta_{\Delta})\left[\bar{\Delta}'^{2}(1+\delta_{\Delta})-\bar{B}'^{2}(1+\delta_{B})\right].\label{eq:rg}
\end{eqnarray}

\begin{figure*}
\subfloat[\label{fig:-1a}]{\includegraphics[height=5.98cm]{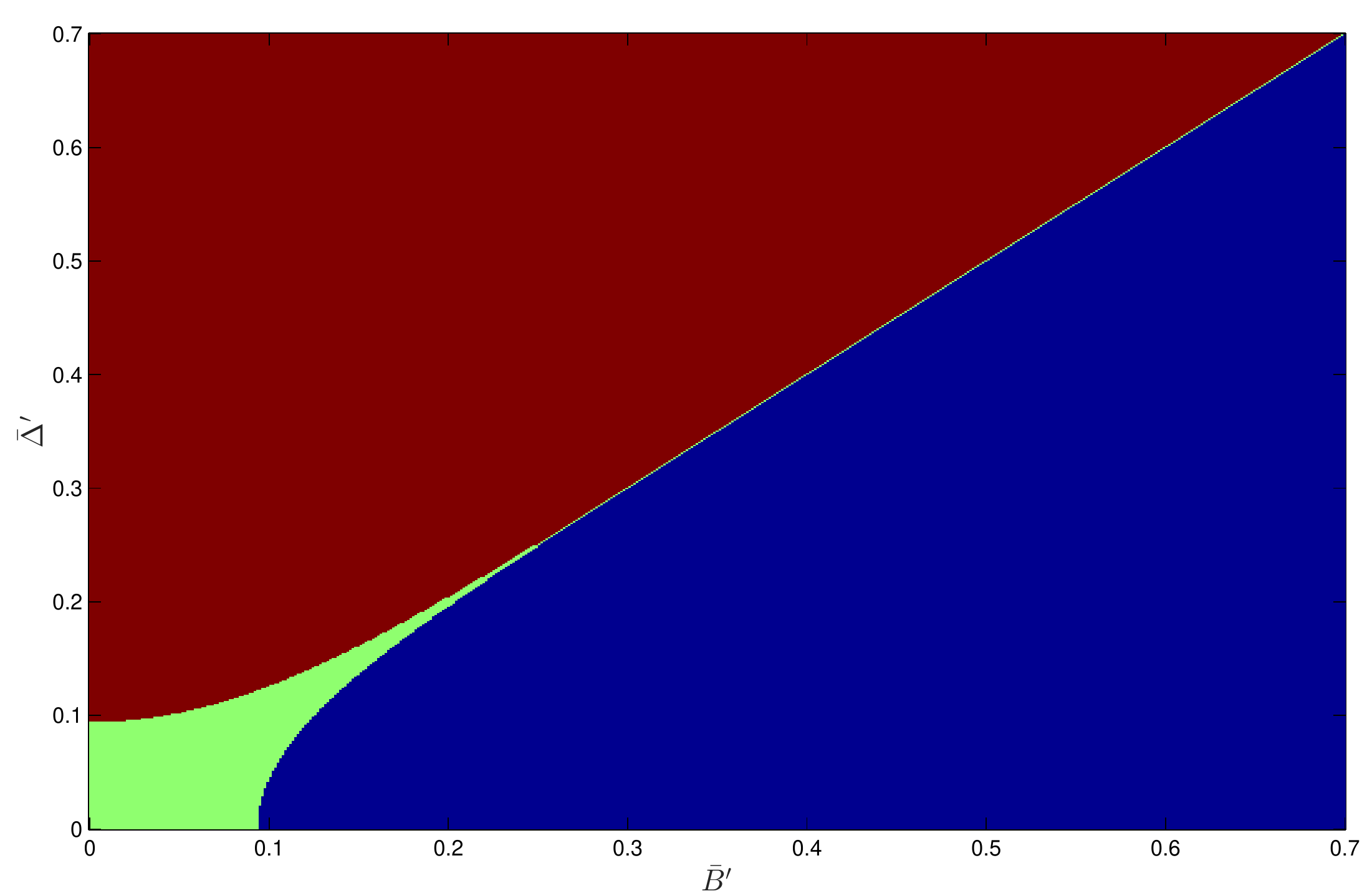}

}\subfloat[\label{fig:-1b}]{\includegraphics[height=6cm]{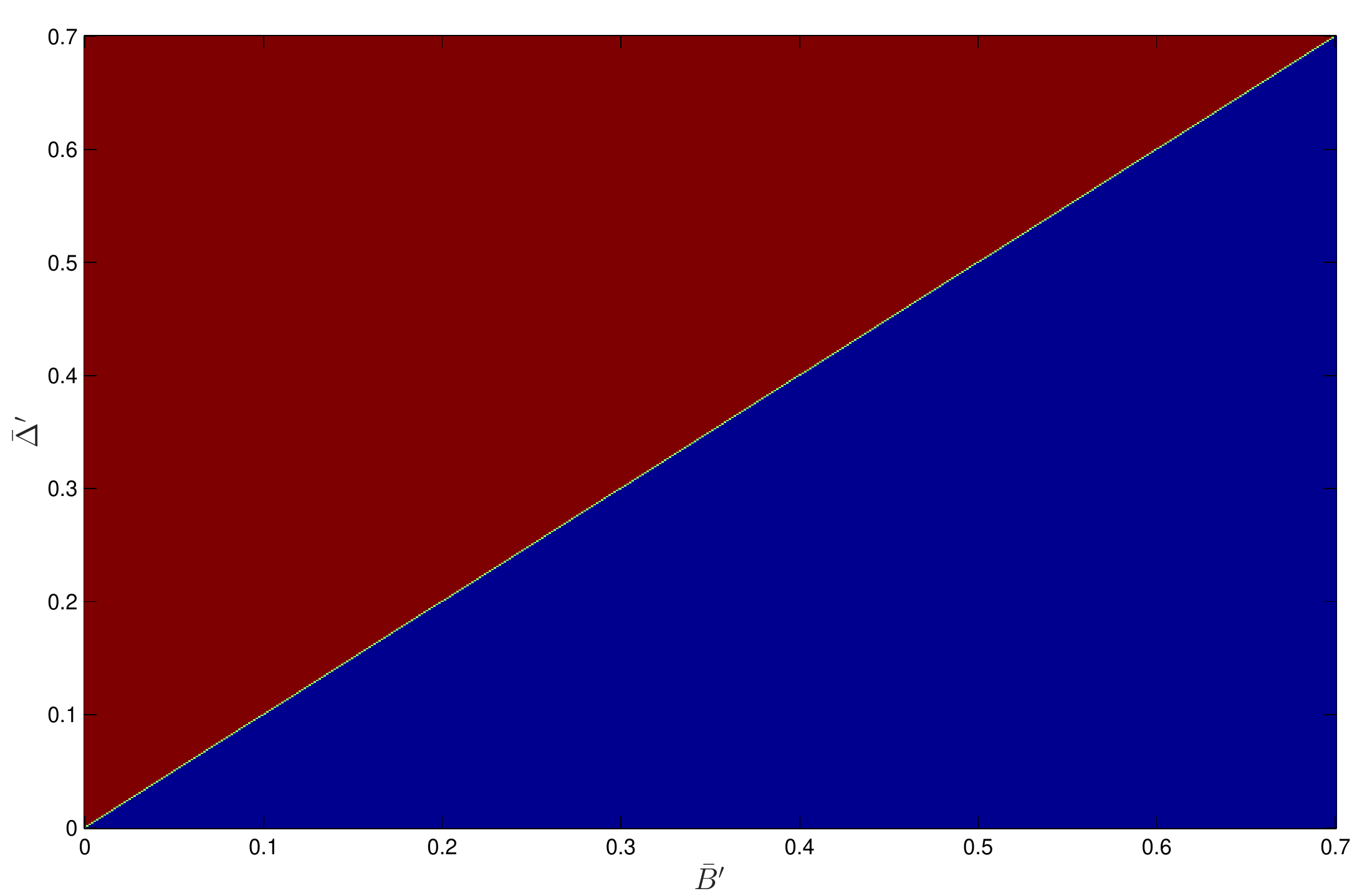}

}

\caption{\label{fig:}The phase diagrams for (a) $\epsilon=0.2$ and (b) $\epsilon=-0.2$.
The red region corresponds to a $\Delta$-dominated phase, the blue
region corresponds to a $B$-terms phase, and the green regions correspond
to gapless states. }
\end{figure*}

Fig. \ref{fig:} presents the phase diagram resulting from the solution
of the RG equations shown in Eq. \ref{eq:rg}. Specifically, the different
colors represent different phases based on the flow equations: The
red regions represent a $\Delta$-dominated phase, the blue region
represents a $B$-dominated phase, and the green regions represent
a gapless state. Figure \ref{fig:-1a} (\ref{fig:-1b}) were generated
by studying the properties of the RG equations for various initial
values of $\bar{B}',\bar{\Delta'}$ and $\delta_{B}=\delta_{\Delta}=\frac{\epsilon}{2}$,
with $\epsilon=0.2$ ($\epsilon=-0.2$) .

\subsection{Studying the self-dual line}

Focusing on the self-dual line $\bar{B}'=\bar{\Delta}'=\bar{\lambda}$,
$\delta_{B}=\delta_{\Delta}=\epsilon/2$, we get a single RG equation

\[
\frac{d\bar{\lambda}}{dl}=-\epsilon\bar{\lambda}+\bar{\lambda}^{3},
\]
accompanied by $\frac{d\epsilon}{dl}=0$.

We immediately find a multicritical point at $\bar{\lambda}=\sqrt{\epsilon}$,
above which a flow to large $\bar{\lambda}$ ensues.

This shows that if the coupling constants in the self-dual theory
studied in the main text are large enough, the cosine terms
flow to large coupling.

As we argued in the main text, if $\bar{\lambda}$ is tuned to the multicritical point, the low energy field
theory is not given by the original Luttinger liquid fixed point,
but rather by a parafermion CFT.

\section{The scaling dimension of the local electron fields}\label{appendix:scaling}

In this section, we demonstrate that the scaling dimension of the
local electron fields $\tilde{\alpha}_{nR}(x),\tilde{\beta}_{nL}(x)$
remains $m/2$ in the presence of the intra-wire interacting term,
which is taken to the critical point $\tilde{\Delta}=\tilde{B}=\lambda$:
\begin{equation}
\tilde{H}_{\Delta}+\tilde{H}_{B}=\lambda i\int dx\tilde{\beta}_{nR}\tilde{\alpha}_{nL}.\label{eq:intra-wire}
\end{equation}
We first note that since $\tilde{\alpha}_{nR}(x)$ and $\tilde{\beta}_{nL}(x)$
commute with the above interacting Hamiltonian, they remain right
and left moving fields, respectively. To demonstrate this, we write
the equations of motion of $\tilde{\alpha}_{nR}(x)$:
\[
\frac{\partial}{\partial\tau}\tilde{\alpha}_{nR}(x,\tau)=\left[H,\tilde{\alpha}_{nR}(x,\tau)\right].
\]
The Hamiltonian is composed of the kinetic chiral Luttinger liquid
Hamiltonian and the interacting Hamiltonian written in Eq. \ref{eq:intra-wire}.
Since $\tilde{\alpha}_{nR}(x)$ commutes with the latter, we only
need to calculate the commutation with the kinetic part,
\[
H_{K}=\frac{v}{4\pi m}\lim_{\epsilon\rightarrow0}\int\left\{ \partial\eta_{nR}(x)\right\} \left\{ \partial\eta_{nR}(x+\epsilon)\right\} ,
\]
where we have used point splitting regularization. Since, by definition,
$\tilde{\alpha}_{nR}(x)=e^{i\eta_{nR}}+e^{-i\eta_{nR}}$, we calculate
the commutation relation with each vertex operator separately:
\begin{widetext}
\begin{align*}
\left[H,e^{\pm i\eta_{nR}(x)}\right] & =\frac{v}{4\pi m}\lim_{\epsilon\rightarrow0}\int dx'\left[\left\{ \partial\eta_{nR}(x')\right\} \left\{ \partial\eta_{nR}(x'+\epsilon)\right\} e^{\pm i\eta_{nR}(x)}-e^{\pm i\eta_{nR}(x)}\left\{ \partial\eta_{nR}(x')\right\} \left\{ \partial\eta_{nR}(x'+\epsilon)\right\} \right]\\
 & =e^{\pm i\eta_{nR}(x)}\frac{v}{4\pi m}\lim_{\epsilon\rightarrow0}\int dx'\left[e^{\mp i\eta_{n,R}}\left\{ \partial\eta_{nR}(x')\right\} \left\{ \partial\eta_{nR}(x'+\epsilon)\right\} e^{\pm i\eta_{nR}(x)}-\left\{ \partial\eta_{nR}(x')\right\} \left\{ \partial\eta_{nR}(x'+\epsilon)\right\} \right].
\end{align*}
\end{widetext}
Using the commutation relations
\[
\left[\eta_{nR}(x),\partial\eta_{nR}(x')\right]=-2\pi im\delta(x-x')
\]
and the Baker-Campbell-Hausdorff formula, we write
\begin{align*}
 & e^{\mp i\eta_{n,R}}\left\{ \partial\eta_{nR}(x')\right\} e^{\pm i\eta_{nR}(x)}\\
 & =\partial\eta_{nR}(x')\mp i\left[\eta_{nR}(x),\partial\eta_{nR}(x')\right]\\
 & =\partial\eta_{nR}(x')\mp2\pi m\delta(x-x'),
\end{align*}
and therefore
\[
\left[H,e^{\pm i\eta_{nR}(x)}\right]=\mp ve^{\pm i\eta_{nR}(x)}\partial\eta_{nR}(x)=-iv\partial e^{\pm i\eta_{nR}(x)}.
\]
This gives us the equation of motion
\[
\frac{\partial}{\partial\tau}\tilde{\alpha}_{nR}(x,\tau)=-iv\frac{\partial}{\partial x}\tilde{\alpha}_{nR}(x,\tau).
\]
Defining $z=x+iv\tau,$ we may write the above as
\[
\frac{\partial}{\partial\bar{z}}\tilde{\alpha}_{nR}(x,\tau)=0.
\]

This shows that $\tilde{\alpha}_{nR}(z)$ remains a right mover in
the presence of the interacting Hamiltonian \ref{eq:intra-wire}.
Similarly, $\tilde{\beta}_{nL}$ satisfies $\frac{\partial}{\partial z}\tilde{\beta}_{nL}(x,\tau)=0$,
and is therefore a left mover. We note that the above analysis cannot
be repeated for the fields $\tilde{\alpha}_{nL}$ and $\tilde{\beta}_{nR}$
which do not commute with the interacting Hamiltonian, and are therefore
not chiral.

Using the above result, we now turn to calculate the propagator of
the remaining chiral fields and show that all corrections to the $\lambda=0$
limit vanish identically. For example, calculating the propagator
$g(z-z')=\left\langle \tilde{\alpha}_{nR}(z)\tilde{\alpha}_{nR}(z')\right\rangle $,
and treating the interacting Hamiltonian shown in Eq. \ref{eq:intra-wire}
perturbatively, we can write the correction to any order $p$ as
\begin{align}
 & \delta g_{p}(z-z')\propto\nonumber \\
 & \left(\lambda i\right)^{p}\left\langle \tilde{\alpha}_{nR}(z)\tilde{\alpha}_{nR}(z')\Pi_{i=1}^{p}\int d^{2}z_{i}\tilde{\beta}_{nR}(z_{i},\bar{z}_{i})\tilde{\alpha}_{nL}(z_{i},\bar{z}_{i})\right\rangle _{0}.\label{eq:higher order correlation}
\end{align}

Since the expectation value is done with respect to the Kinetic Hamiltonian,
we can treat $\tilde{\alpha}_{nL}$ as a function $\bar{z}$ only
and $\tilde{\beta}_{nR}$ as a function of $z$. Furthermore, we can
use the symmetries of the Kinetic Hamiltonian: First, we know that
under rotations, $z\rightarrow ze^{i\theta},\bar{z}\rightarrow\bar{z}e^{-i\theta}$,
the fields transform as $\tilde{\alpha}_{nR}(z)\rightarrow\tilde{\alpha}_{nR}(z)e^{-i\theta m/2},\tilde{\alpha}_{nL}(\bar{z})\rightarrow\tilde{\alpha}_{nL}(\bar{z})e^{i\theta m/2}$
(and the same relations for the $\beta$-fields). This leads to the
condition
\[
\delta g_{p}\left[\left(z-z'\right)e^{i\theta}\right]=e^{-i\theta m}\delta g_{p}\left[z-z'\right],
\]
which is satisfied if $\delta g_{p}\propto(z-z')^{-m}.$

Under scale transformations, given by $z\rightarrow zd,\bar{z}\rightarrow\bar{z}d$,
the fields transform as $\tilde{\alpha}_{nR}(z)\rightarrow\tilde{\alpha}_{nR}(z)d^{-m/2},\tilde{\alpha}_{nL}(\bar{z})\rightarrow\tilde{\alpha}_{nL}(\bar{z})d^{-m/2}$
(and similarly for the $\tilde{\beta}$-fields). In this case we get
the condition $\delta g_{p}\left[\left(z-z'\right)d\right]=d^{-m+p(2-m)}\delta g_{p}\left[z-z'\right],$
which is satisfied if $\delta g_{p}\propto(z-z')^{-m+p(2-m)}.$ Clearly,
we have reached a contradiction, which can only be solved if $\delta g_{p}\left[z-z'\right]=0$
for any $p>0$. This shows that all corrections to the propagator
calculated in the absence of the interacting term vanish, and the
scaling dimension of $\tilde{\alpha}_{nR}$ and $\tilde{\beta}_{nL}$
therefore remains $m/2$.

We note that as the fields $\tilde{\beta}_{nR}$ and $\tilde{\alpha}_{nL}$
are no longer chiral, the full propagator is not a holomorphic (or
antiholomorphic) function, and the entire argument fails, as expected.

\end{appendix}

\end{document}